\documentclass{aa}  
\usepackage{graphicx}
\usepackage{txfonts}
\usepackage{natbib}
\bibpunct{(}{)}{;}{a}{}{,} 

\newcommand{\wda}{WD~1310+583}

\usepackage[normalem]{ulem}
\usepackage{color}

\usepackage{hyperref}
\hypersetup{
    colorlinks = true,
    citecolor ={blue}
}

\usepackage{placeins}

\begin{document}

   \title{Asteroseismology of the ZZ~Ceti star WD~1310+583 using the Transiting Exoplanet Survey Satellite}


    \author{Zs\'ofia~Bogn\'ar\inst{1}\thanks{\email{bognar.zsofia@csfk.org}},
            Murat~Uzundag\inst{2},
            Francisco C. De Ger\'onimo\inst{3,4}, 
            Alejandro H. C\'orsico\inst{3,4},
            James Munday\inst{5},
            \'Ad\'am~S\'odor\inst{1},
            Sam D. Barber\inst{6,7}
            }

   \institute{
        Konkoly Observatory, HUN-REN Research Centre for Astronomy and Earth Sciences, MTA Centre of Excellence, H-1121 Budapest, Konkoly Thege Mikl\'os \'ut 15-17, Hungary
        \and
        Institute of Astronomy, KU Leuven, Celestijnenlaan 200D, B-3001 Leuven, Belgium
        \and
        Instituto de Astrofísica de La Plata, IALP (CCT La Plata), CONICET-UNLP, Argentina
        \and
         Grupo de Evolución Estelar y Pulsaciones. Facultad de Ciencias Astronómicas y Geofísicas, Universidad Nacional de La Plata, Paseo del Bosque s/n, (1900) La Plata, Argentina
        \and
        Department of Physics, Gibbet Hill Road, University of Warwick, Coventry CV4 7AL, United Kingdom
        \and
        Department of Astronomy, University of Texas at Austin, Austin, TX-78712, USA
        \and
        McDonald Observatory, Fort Davis, TX-79734, USA
        }
        
    \titlerunning{TESS observations of \wda}
        \authorrunning{Zs.~Bogn\'ar et al.}
        
    \date{}

 
  \abstract
   {}
   {By analysing the light curves of the ZZ~Ceti star \wda, we aim to determine its pulsational frequencies and to give constraints on the main stellar parameters using the tools of asteroseismology.}
   {We performed the Fourier analysis of the TESS light curves of \wda\ and selected the possible pulsational modes. We also used spectroscopic data collected with the Cosmic Origins Spectrograph of the Hubble Space Telescope to give constraints for the asteroseismic analysis. We perform the latter with period-to-period fits using fully evolutionary white dwarf models.}
   {The star presented in this paper shows a particularly large number (41) of pulsational frequencies, which provides a potential 
   opportunity for detailed asteroseismic investigations. We found a mean period spacing of $\sim 40.5$ seconds, which allows us to state that the stellar mass of WD~1310+583 would be larger than $\sim 0.57 M_{\odot}$. We also attempted an asteroseismological analysis by performing period-to-period fits, but we were unable to find a single statistically significant asteroseismological solution. We adopted a tentative solution consisting of a white dwarf model with $M_*=0.632\,M_{\odot}$, $T_{\rm eff}= 11\,702\,$K, and an  asteroseismic distance $d = 27.75^{+0.17}_{-0.15}$\,pc, which is significantly smaller than the one predicted by Gaia ($d = 30.79\pm 0.2$\,pc). 
   We also determined that the rotational period of our target is 1.18\,d.}
   {}


   \maketitle
%

\section{Introduction}

ZZ~Ceti stars, also known as DAV stars, are pulsating white dwarfs with hydrogen-dominated atmospheres. These stars exhibit low-amplitude multiperiodic brightness variations caused by nonradial \textit{g}-mode oscillations with periods ranging from 100 to 1500 seconds. These pulsations are driven by a combination of the $\kappa-\gamma$ mechanism, operating within the partial ionisation zone of hydrogen \citep{1981A&A...102..375D, 1982ApJ...252L..65W}, and the convective driving mechanism \citep{1991MNRAS.251..673B, 1999ApJ...511..904G}. With effective temperatures spanning 10~500 to 13~000\,K, ZZ~Ceti stars are confined to a well-defined region in the Hertzsprung--Russell diagram known as the ZZ~Ceti instability strip.

As these stars cool across the instability strip, their pulsational properties vary systematically. Near the blue (hot) edge, they exhibit a smaller number of pulsation modes with lower amplitudes and shorter periods. Toward the red (cool) edge, the number of detectable modes increases, their amplitudes grow, and the pulsations become less stable, often showing amplitude and frequency variations on timescales of days to weeks (see, e.g., section~6 in \citealt{2008PASP..120.1043F}). Intriguingly, many cool ZZ~Ceti stars also display episodic outbursts, characterised by brief and irregular increases in stellar brightness (see \citealt{2015ApJ...809...14B, 2016ApJ...829...82B, 2017ASPC..509..303B, 2023A&A...674A.204B} and \citealt{2015ApJ...810L...5H}), which may provide insight into the interaction between pulsation and convection.

ZZ~Ceti stars are invaluable astrophysical laboratories for investigating the physics of dense matter under extreme conditions. Their small size (approximately Earth-like) and high surface gravity ($\log g \sim 8$) result in strongly stratified atmospheres, where heavier elements settle below a hydrogen or helium layer due to gravitational separation. These characteristics make them ideal for asteroseismology, a technique that uses observed pulsation periods to infer the star’s internal structure, chemical composition, and rotational properties. Such studies contribute to understanding stellar evolution and the formation of compact objects \citep{2008ARA&A..46..157W, 2008PASP..120.1043F, 2010A&ARv..18..471A, 2019A&ARv..27....7C, 2020FrASS...7...47C}.

Recent advances in space-based photometry, particularly with the Transiting Exoplanet Survey Satellite (TESS; \citealt{2015JATIS...1a4003R}), have revolutionised the study of ZZ~Ceti stars. TESS provides high-precision, high-cadence observations, enabling detailed frequency analyses and the detection of new pulsation modes, even in stars previously deemed observationally challenging. This has opened new avenues for understanding the global properties of white dwarfs and the underlying physics driving their pulsations \citep{2019A&A...632A..42B, 2020A&A...638A..82B}.

The brightness variations of the ZZ~Ceti star \wda\ were independently discovered by \citet{2018MNRAS.473.3693G} and \citet{2018MNRAS.478.2676B}. The latter study presents photometric investigations conducted at the Konkoly Observatory (Hungary). Seven independent modes were identified on the basis of the data sets. In the study by \citet{2018MNRAS.473.3693G}, two possible pulsation frequencies were published as a result of their analysis. The star has been independently suggested as a double degenerate source based on spectroscopic or hybrid (photometric, spectroscopic plus astrometric data) fits using ultraviolet \citep{2018MNRAS.473.3693G} and optical \citep{Munday2024} spectroscopy. \citet{Munday2024} found little radial velocity variability, indicating that the source is likely a wide double white dwarf binary star system.

In this paper, we focus on the TESS measurements of \wda. First, we analysed the spectroscopic data available on \wda, then we presented the TESS data sets to identify the pulsation modes of this star. Using these findings, we estimate the period spacing to constrain the stellar mass of \wda\ and attempt an asteroseismic investigation to constrain its global parameters and internal structure.

\section{Spectroscopic analysis}
\label{Joint_fit}

In the following, we present the results of the system parameters of a double-degenerate fit to the spectra obtained on \wda\ (TIC\,157271533, $G = 14.07$\,mag, $\alpha_{2000}=13^{\mathrm h}12^{\mathrm m}58^{\mathrm s}$, $\delta_{2000}=+58^{\mathrm d}05^{\mathrm m}11^{\mathrm s}$).

At the time of writing, precise Gaia parallax measurements were not available to \citet{2018MNRAS.473.3693G} and the authors chose to fix the surface gravity of both stars to $\log g=8.0$\,dex when producing atmospheric parameters of the two stars. With that in mind and to obtain an improved spectroscopic solution to the ultra-violet data, we decided to refit their Hubble Space Telescope Cosmic Origins Spectrograph (HST COS) spectrum with new information at hand, utilising all-sky photometry and Gaia parallaxes to give absolute flux measurements. 

We used the WD-BASS pipeline \citep{Munday2024} with Pan-STARRS all-sky photometry \citep[][filters \textit{grizy}]{Panstarrs}, the Sloan Digital Sky Survey (filters \textit{ugriz}) DR16 \citep{SDSSdr16}, and the Gaia DR3 parallax of $\pi=32.48\pm0.25$\,mas. We fit the wavelength range of 1200--1930\,\AA, trimming the very low signal-to-noise data. We mask geocoronal lines between vacuum wavelengths 1206--1226\,\AA, 1295--1315\,\AA~and also within 1\,\AA~of all entries supplied in the line list of \citet{Sahu2023} to ignore any potential undesired photospheric/interstellar flux contribution.

Synthetic spectra were obtained by interpolating the 3D local thermodynamic equilibrium grids of \citet{Tremblay2013, Tremblay2015} for DA white dwarfs, which are based on the line profiles of \citet{Tremblay2009}. The hotter star, which dominates the flux, has Balmer absorption lines and is clearly a DA, but the spectral class of the secondary star is unknown since it exhibits no unique spectral lines. Hence, the dimmer companion could be a helium-atmosphere DC, and in testing this DA+DC combination, we interpolate synthetic spectra from \citet{ElenaCukanovaite2021_3D_DB}. The temperature, surface gravity, and radial velocity of the two stars and the parallax of the system were independent variables in the fitting. We used the mass–temperature–radius relationships from the hydrogen-rich (DA) and helium-rich (DC) evolutionary sequences of \citet{Bedard2020} to obtain stellar radii and calculate the luminosity of each star.
Observations were scaled from an Eddington flux to an observed flux using the reciprocal of the parallax as the distance in parsecs, with an extinction coefficient of $\text{E(B}-\text{V})=\text{A}_V/\text{R}_V=0.01/3.1$ applied to redden the synthetic spectra \citep{2021MNRAS.508.3877G}. No spectrum normalisation is applied; the observed flux in physical units is fitted in both the HST COS spectrum and the all-sky photometry with a common parallax measurement to each dataset.  A Gaussian prior was placed on the fitted parallax using the parallax, and the error reported in Gaia DR3.

However, as it turns out, we need to be careful with the automatic use of the Gaia parallax. The value of the parameter of the renormalised unit weight error (RUWE) for WD\,1310+583 (Gaia DR3 1566603962760532736) is quite high; 16.315. This may also indicate the presence of a binary star system, as the RUWE parameter measures the goodness of an astrometric fit to the data (see, e.g. \citealt{2024A&A...688A...1C}). We checked the Gaia DR3 database \citep{2022yCat.1355....0G} and found that the parameter \texttt{astrometric\_excess\_noise} is 2.688\,mas, the \texttt{astrometric\_excess\_noise\_sig} is 4\,076, and the \texttt{visibility\_periods\_used} is 28. It strongly indicates that the `single-star' astrometric model is not adequate: the high RUWE and, in particular, the large \texttt{astrometric\_excess\_noise} together with its extreme significance show that the assumed single-star model fails to explain many of the measurements. The most likely physical cause is photocentre motion (e.g. due to a close binary, an unresolved companion, or blending) or strong photometric variability that shifts the astrometric positions. Thus, the parallax may be inaccurate or partially biassed (the formal \texttt{parallax\_error} is likely too optimistic). We must bear in mind that it is not recommended to use the value of 32.48\,mas automatically as an accurate distance estimate; however, this is currently the best known estimate of the star's distance. The \texttt{visibility\_periods\_used = 28} shows that there was a sufficient number of observations, so the poor fit is not simply due to a small data sample; the deviation is real.

\begin{figure*}
    \centering
    \includegraphics[width=0.48\textwidth]{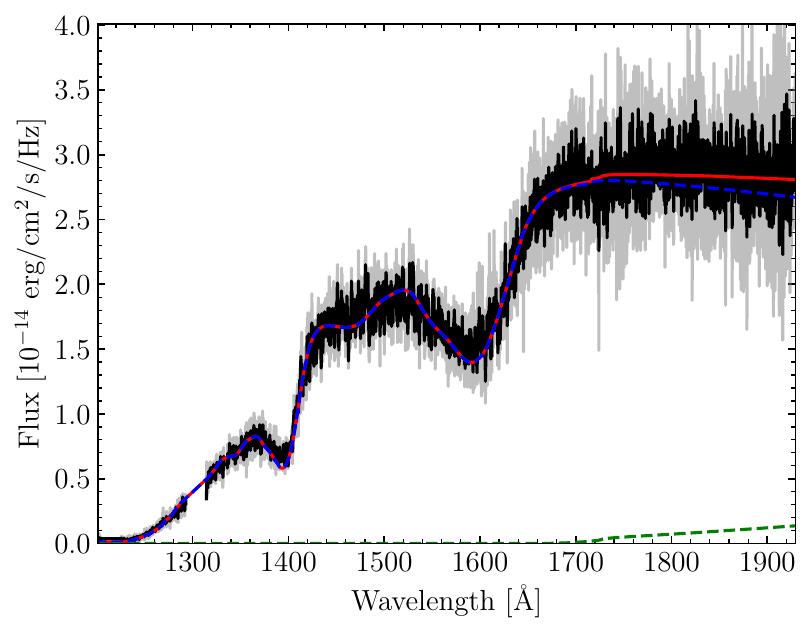}
    \includegraphics[width=0.48\textwidth]{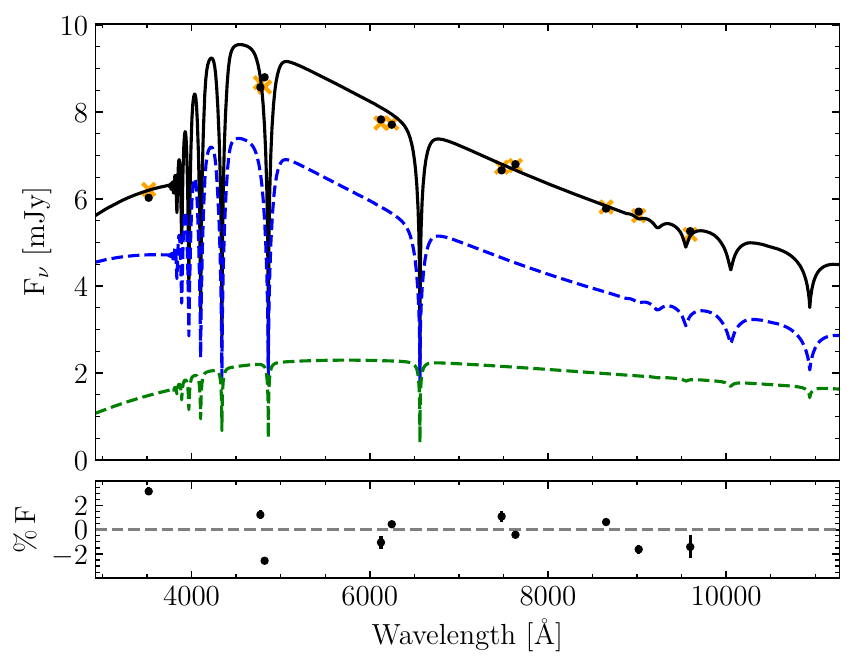}
    \caption{The best-fitting solution for a DA+DA two-star model. The left panel shows the fit to the HST~COS spectrum and the right panel shows the solution with the photometric data. The synthetic flux from the hotter, pulsating white dwarf is shown in dashed blue while the flux from the cooler companion is shown in dashed green. In the left panel, the reduced spectrum is in grey and in black is the reduced spectrum smoothened across 5 data points. In the right panel, the synthetic flux in each filter is given as black circles and the orange crosses are observed fluxes from the photometric surveys, with the percentage flux residual underneath. The total flux is in red on the left and black on the right for clarity.}
    \label{fig:AtmosphericSolutionDADA}
\end{figure*}

\begin{figure*}
    \centering
    \includegraphics[width=0.48\textwidth]{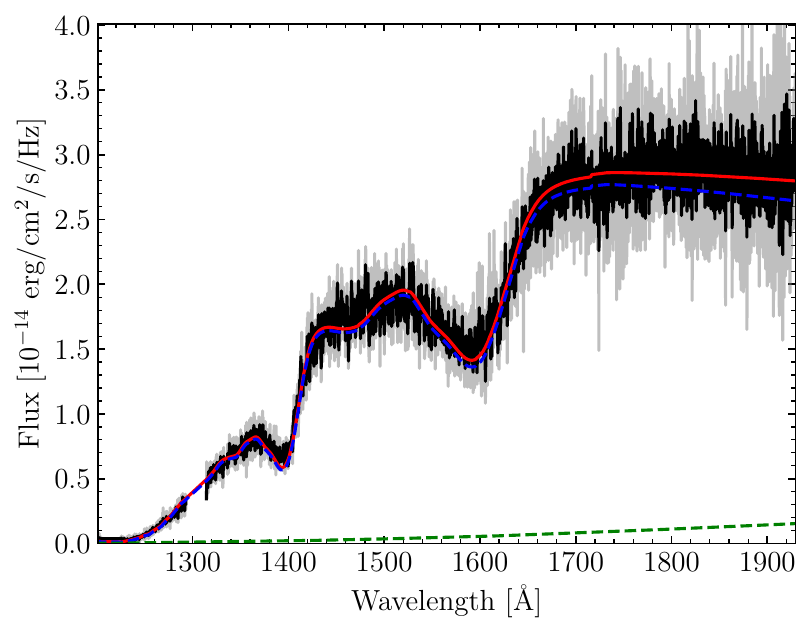}
    \includegraphics[width=0.48\textwidth]{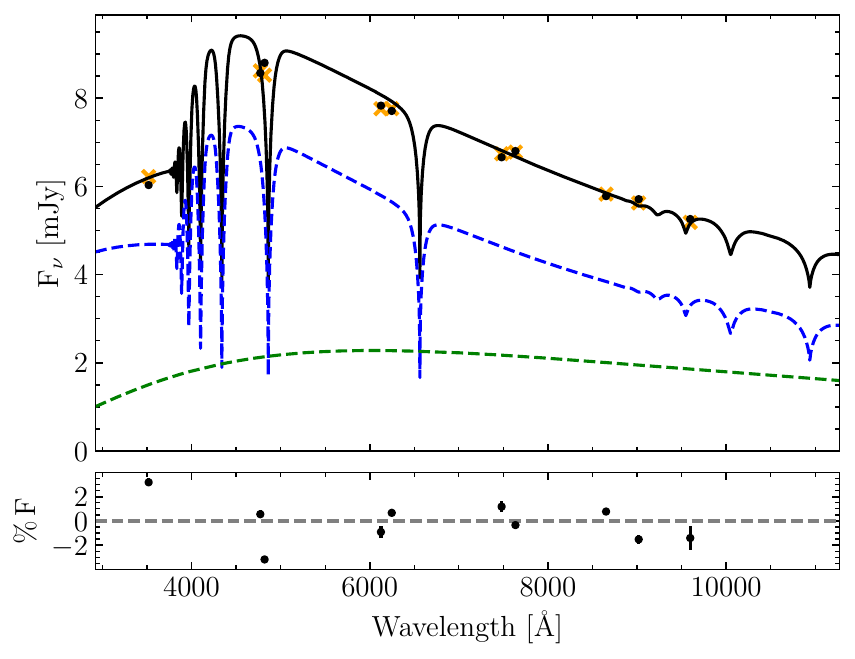}
    \caption{The same as Fig~\ref{fig:AtmosphericSolutionDADA} but for a DA+DC two-star model. See Sect.~\ref{Joint_fit} for more details}.
    \label{fig:AtmosphericSolutionDADC}
\end{figure*}

The best-fit solutions are shown in Fig.~\ref{fig:AtmosphericSolutionDADA} and Fig.~\ref{fig:AtmosphericSolutionDADC}, respectively, while the physical parameters of the different solutions can be seen in Table~\ref{tab:spec}. Like in \citet{2018MNRAS.473.3693G}, we conclude that a single star solution does not suffice to model WD~1310+583. The flux contributed by the companion in the ultra-violet is small, but the relative flux contribution from the companion becomes significant towards redder wavelengths. As such, the companion star serves as an extra flux contributor and there is no way to decisively reveal its spectral type. Both combination of DA+DA and DA+hydrogen-deficient DC is presented for this reason. Crucially, since the flux in the ultra-violet almost completely originates from the hotter object, the atmospheric parameters of the pulsating white dwarf are near identical across the two models, meaning that the unknown combination of DA+DA or DA+DC does not noticeably impact our astroseismetric analysis.

\begin{table}[]
  \centering
\caption{Physical parameters of the different spectroscopic fits.}
\begin{tabular}{lr}
\hline
\hline
\multicolumn{2}{l}{DA+DA fit}\\
T$_1=$ & $11670\pm163$\,K\\
$\log g_1 =$ & $8.13\pm0.04$\,dex\\
T$_2=$ & $8240\pm115$\,K\\
$\log g_2 =$ & $8.09\pm0.04$\,dex\\
Fitted Parallax = & $31.75\pm0.17$\,mas\\
 & \\
\multicolumn{2}{l}{DA+DC fit}\\
T$_1=$ & $11650\pm163$\,K\\
$\log g_1 =$ & $8.12\pm0.04$\,dex\\
T$_2=$ & $7540\pm106$\,K\\
$\log g_2 =$ & $7.78\pm0.04$\,dex\\
Fitted Parallax = & $31.56\pm0.17$\,mas\\
\hline
\end{tabular}
\label{tab:spec}
\tablefoot{For more reflective errors including systematics, we presented an external errors of 1.4\% for $T_{\rm eff}$ and 0.042\,dex for $\log g$, respectively \citep{2005ApJS..156...47L}.}
\end{table}

\section{Light curve analyses of the TESS observations}
\label{sect:obs}

TESS measured the star in 120-second short-cadence mode in these sectors: 15, 16, 22, 48, 49, 75, 76. Measurements are also available in 20-second ultra-short cadence mode, but we did not see peaks above the Nyquist frequency of the 120-second measurements. So, we examine the 120-second measurements more closely.

First, we performed frequency analyses for sectors s15s16, s22, s48s49, and s75s76 using the photometric module of the Frequency Analysis and Mode Identification for Asteroseismology (\textsc{famias}) software package \citep{2008CoAst.155...17Z}. As seems, we treat the data of the neighbouring sectors jointly. For frequency analyses, we set the significance limit at 0.1\% False Alarm Probability (FAP). We created a table of the observed peaks in the different sectors and considered the peaks that are approximately at the same frequency in the different data sets as the same frequency. The full list of detected frequencies can be found in the tables of Appendix~\ref{app:B}. Then, after we identified the common frequencies and their amplitudes provided by the analyses, we calculated the amplitude-averaged period value for each frequency.
A total of 67 frequencies were detected in the individual datasets. Based on pairwise frequency separations, 39 pairs with spacings smaller than $1.22\,\mu$Hz were identified. Since all remaining pairs are separated by at least $2.70\,\mu$Hz, we adopted $1.22\,\mu$Hz as the threshold, below which the frequencies were combined by means of a weighted average.

We also checked for the presence of linear combinations in each data set.

There are groups of peaks above about 700~s, and we considered the one with the highest amplitude as the representative frequency of the given frequency group.

Table~\ref{tabl:journal} summarises the journal of observations, while Appendix~\ref{app:A} lists the frequencies, periods, and amplitudes of the peaks detected by the analysis of the data sets from the different sectors.

\begin{table*}
\centering
\caption{Journal of observations of \wda.}
\label{tabl:journal}
\begin{tabular}{lccrrcrc}
\hline
\hline
Object & TIC ID & Start time & \multicolumn{1}{c}{\textit{N}} & \multicolumn{1}{c}{$\delta T$} & \textit{G} mag & Sect. (Cadence) & \multicolumn{1}{c}{CROWDSAP} \\
& & (BJD-2\,457\,000) & & \multicolumn{1}{c}{(d)} & & & \\
\hline
\wda\ & 157271533 & 1711.362 &  29\,058 & 51.2 & 14.1 & 15--16 (120\,s) & 0.99 \\
 &  & 1900.094 & 16\,326 & 26.4 & & 22 (120\,s) & \\
 &  & 2610.234 & 28\,155 & 54.1 & & 48--49 (120\,s) & \\
 &  & 3339.788 & 21\,531 & 55.0 & & 75--76  (120\,s) & \\
\hline
\end{tabular}
\tablefoot{\textit{TIC ID} refers to the TESS Input Catalog identifier of the object, \textit{N} is the number of data points after cleaning the light curve, $\delta T$ is the total length of the data sets including gaps, and \textit{Sect.} is the serial number of the sector(s) in which the star was observed. The start time in BJD is the time of the first data point in the data set. The \textit{CROWDSAP} keyword represents the ratio of the target flux to the total flux in the TESS aperture. We note that here we list the CROWDSAP value from the first TESS run.}
\end{table*}

\begin{figure*}
\centering
\includegraphics[width=0.90\textwidth]{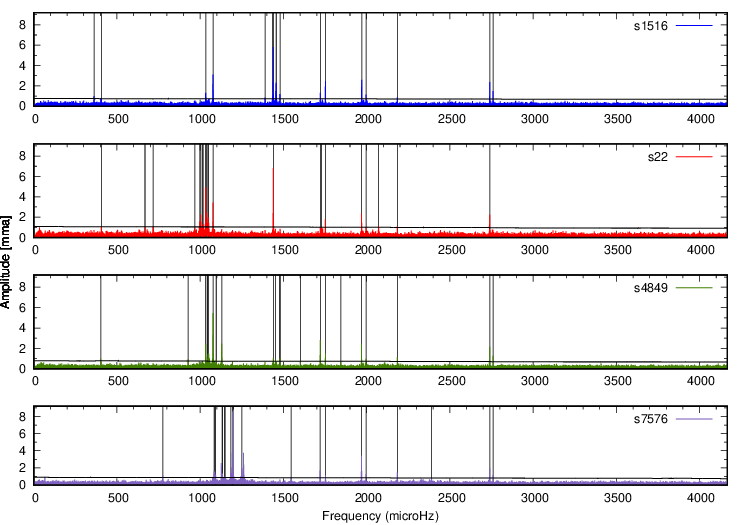}
\caption{Fourier spectra of the different light curve segments. Black horizontal lines indicate the significance levels corresponding to a 0.1\% FAP for each dataset. Amplitude variations are clearly visible from sectors to sectors. The vertical lines in each panel correspond to the frequencies listed in Appendix~\ref{app:B}.}
\label{fig:sps}
\end{figure*}

The comparison of the Fourier spectra of the different segments of the light curve is shown in Fig.~\ref{fig:sps}.

\section{Rotational multiplets}

The equation we used to calculate the rotation period of the star is as follows, see e.g. the Appendix in \citet{1991ApJ...378..326W}.
\begin{equation}
\label{eq:rot}
    \delta f_{k,l,m} = m(1-C_{k,l})\Omega , 
\end{equation}
\noindent where $k$, $l$, and $m$ are the radial order, horizontal degree, and azimuthal order of the non-radial pulsation mode, respectively. The coefficient $C_{k,l}$ can be calculated as $C_{k,l} \approx 1/\ell(\ell+1)$. This relation is valid for high-overtone ($k\gg \ell$) $g$-modes. $\Omega$ is the (uniform) rotational frequency.

\begin{figure*}
\centering
\includegraphics[width=0.95\textwidth]{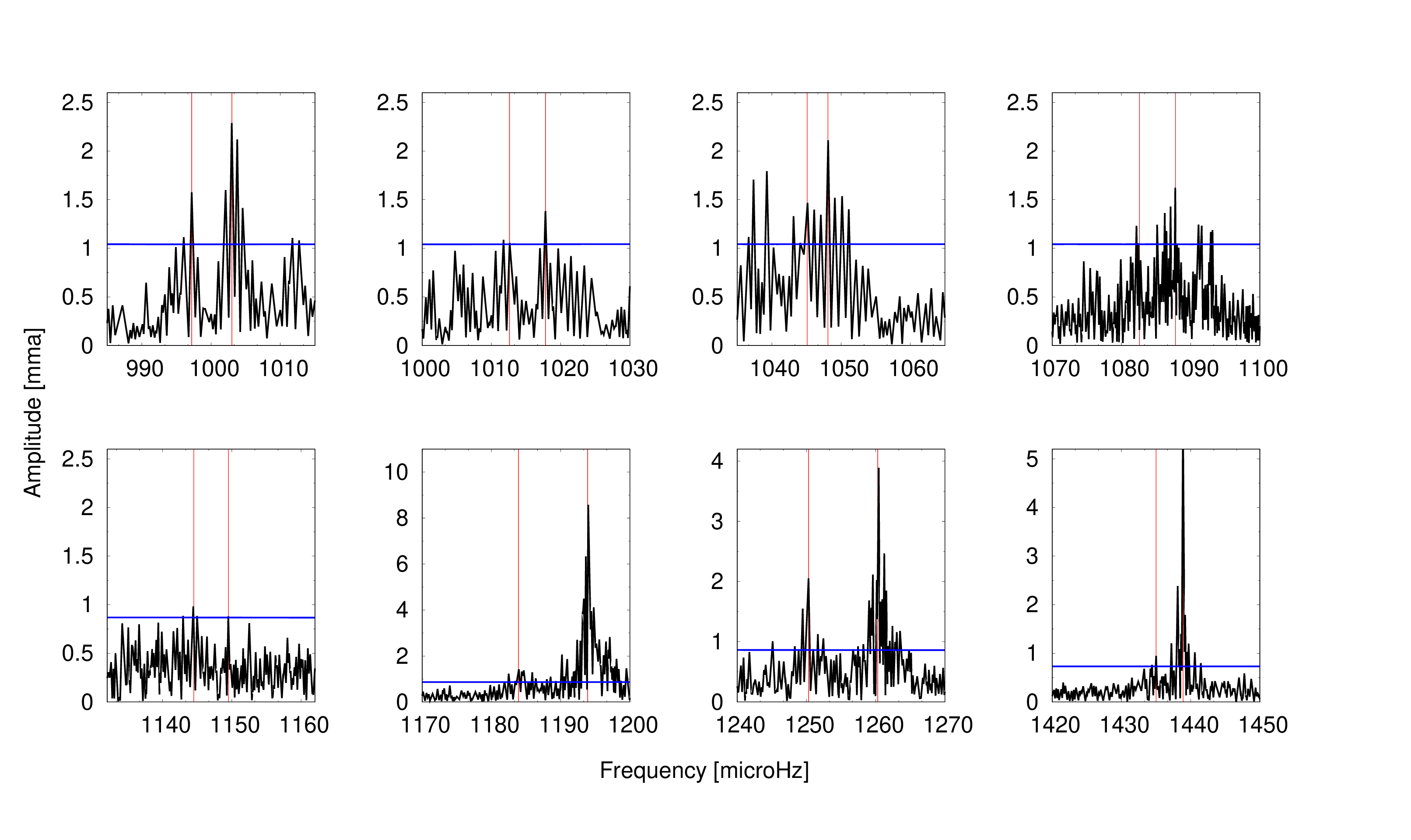}
\caption{Fourier spectra of possible rotational frequencies listed in Table~\ref{tabl:rot}. We note that we utilized pre-whitened Fourier spectra for the plots, as the lower-amplitude peaks become more clearly visible after pre-whitening. Blue horizontal lines indicate the significance levels corresponding to a 0.1\% FAP for each dataset.}
\label{fig:rot}
\end{figure*}

Examining the frequencies listed in the Appendix~\ref{app:A} reveals the presence of possible rotational triplet and doublet frequencies separated by approximately 5 and 10\,$\mu$Hz, respectively. The 10\,$\mu$Hz separations may correspond to triplets with 5\,$\mu$Hz frequency differences. A total of eight doublets were identified, as summarised in Fig.~\ref{fig:rot} and Table~\ref{tabl:rot}. Assuming that all are $l = 1$ modes and that the average frequency separation is 4.9\,$\mu$Hz, the star's rotational period is calculated to be $\mathrm{P} = 1.18$\,d considering Eq.~\ref{eq:rot}. This value is in good agreement with the hour-to-day rotation periods typically observed in pulsating white dwarf stars, see, e.g. Section~7 in \citet{2017ApJS..232...23H}.

\begin{table}[]
  \centering
\caption{Possible rotational doublet frequencies.}
\begin{tabular}{rr}
\hline
\hline
\multicolumn{1}{c}{$f\,[\mu$Hz]} & \multicolumn{1}{c}{$\delta f\,[\mu$Hz]} \\
\hline
997.2 & \\
1003.0 & 5.8 \\
 & \\
1012.6 & \\
1017.8 & 5.2 \\
 & \\
1044.8 & \\
1048.6 & 3.8 \\
 & \\
1082.6 & \\
1087.8 & 5.2 \\
 & \\
1144.5 & \\
1149.5 & 5.0 \\
 & \\
1183.9 & \\
1193.9 & 10.0 \\
 & \\
1250.3 & \\
1260.3 & 10.0 \\
 & \\
1435.0 & \\
1438.9 & 3.9 \\
\hline
\end{tabular}
\label{tabl:rot}
\tablefoot{We also mark the frequency differences in the second column.}
\end{table}

\section{Period spacing tests}
\label{ps_tests}

\begin{figure}
\centering
\includegraphics[width=0.49\textwidth]{tests.eps}
\caption{Results of the inverse variance (I-V, black), Kolmogorov-Smirnov (K-S, blue), and Fourier Transform (F-T, red) statistical tests applied to a subset of 11 periods marked with asterisks in Appendix~\ref{app:A}. The three tests point to the existence of a period spacing of $40.58$ s in \wda, which can be associated to $\ell= 1$ modes. The presence of the subharmonic of this spacing at $\sim 20$ seconds is also apparent.}
\label{fig:tests}
\end{figure}

\begin{figure}
\centering
\includegraphics[width=0.49\textwidth]{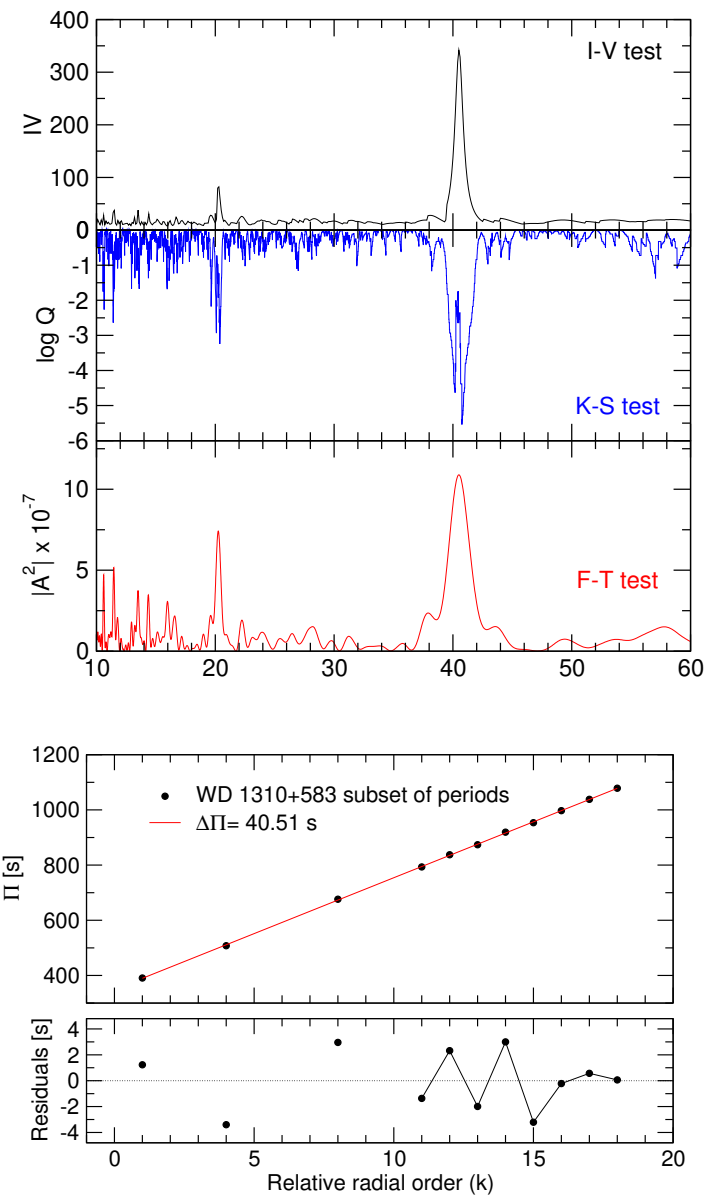}
\caption{Upper panel: linear least-squares fit to the 11 periods of \wda~ marked with asterisks in 
Appendix~\ref{app:A}. The derived period spacing from this fit is $\Delta \Pi_{\ell= 1}= 40.51$\,s. Lower panel: residuals of the period distribution relative to the mean period spacing, revealing signals of mode trapping in the period spectrum of \wda. Modes with a consecutive radial order are connected with thin black lines.}
\label{fig:fit}
\end{figure}

Here, our primary goal is to deepen our understanding of stellar structure, particularly the mass of the observed object. We note that the photometric and spectroscopic masses are in good agreement, as shown in Table~B.1. of the following work: \cite{2024A&A...691A.194C}. The mean period spacing serves as an indicator of the stellar mass. We conducted several tests, as follows: using a subset of the periods listed in Appendix~\ref{app:A}, we searched for a characteristic period spacing using Kolmogorov-Smirnov (K-S; \citealt{1988IAUS..123..329K}) and inverse variance (I-V; \citealt{1994MNRAS.270..222O}) significance tests. In the K-S test, the quantity $Q$ represents the probability that the observed periods are randomly distributed. A characteristic period spacing in the period spectrum would manifest itself as a minimum in $Q$. Meanwhile, in the I-V test, a maximum indicates the presence of a constant period spacing. Another widely recognised approach to identifying a characteristic spacing value is performing a Fourier analysis on a Dirac comb constructed directly with periods (e.g. \citealt{1991ApJ...378..326W,1997MNRAS.286..303H}). 

Fig. \ref{fig:tests} shows the results of applying the statistical tests to a subset of 11 periods detected in \wda. 
These periods are marked with an asterisk in Appendix~\ref{app:A}. The three tests reveal a clear period spacing of $40.58$ s, which can be associated with modes that have $\ell= 1$. The presence of a spacing of $\sim 20$ seconds is also observed in the three tests. This corresponds to the subharmonic of the spacing ($\Delta \Pi /2$). 
Using a linear least-squares fit to the identified dipole modes, we derive a mean period spacing of $\Delta \Pi_{\ell=1} = 40.51$\,s (see Fig. \ref{fig:fit}). To robustly estimate the associated uncertainty, we adopted the error-propagation method described by \citet{2023MNRAS.526.2846U}, which is based on the approach of \citet{2021A&A...651A.121U}. This method involves generating 1\,000 random permutations of the observed periods, in each case assigning a value of $m \in \{-1, 0, +1\}$ for triplets (and $m \in \{-2, -1, 0, +1, +2\}$ for quintuplets) to all detected modes. Then each set is adjusted to the intrinsic $m=0$ component assuming rotational splitting, and a new linear fit is performed. The uncertainty in the mean period spacing is taken as the standard deviation of the resulting distribution of best-fit slopes, yielding $\Delta \Pi_{\ell=1} = 40.51^{+1.99}_{-1.96}$\,s. The residuals of the observed periods relative to this mean period spacing (lower panel of Fig.~\ref{fig:fit}) clearly reveal deviations consistent with mode trapping signatures in the pulsation spectrum of \wda.

The discovery of the period spacing $\Delta \Pi_{\ell= 1}$ allows the harmonic
degree $\ell= 1$ to be assigned to the 11 periods that make up the sequence, allowing to place strong constraints to fits of individual periods (see Section \ref{apf}).

In summary, we first use rotational multiplets to identify the most secure modes. These identifications then serve as priors when we search for the asymptotic period-spacing pattern. We applied this procedure consistently, and the two methods are not in conflict, they are complementary.

\section{Asteroseismology}
\label{seismo}

All of our efforts to determine the independent pulsation modes were for to provide these modes for the asteroseismic analysis of the star presented in the next sections. 

\subsection{The stellar mass of \wda\ as predicted by the observed period spacing}
\label{aps}

\begin{figure*}
	\includegraphics[width=clip,width=500pt]{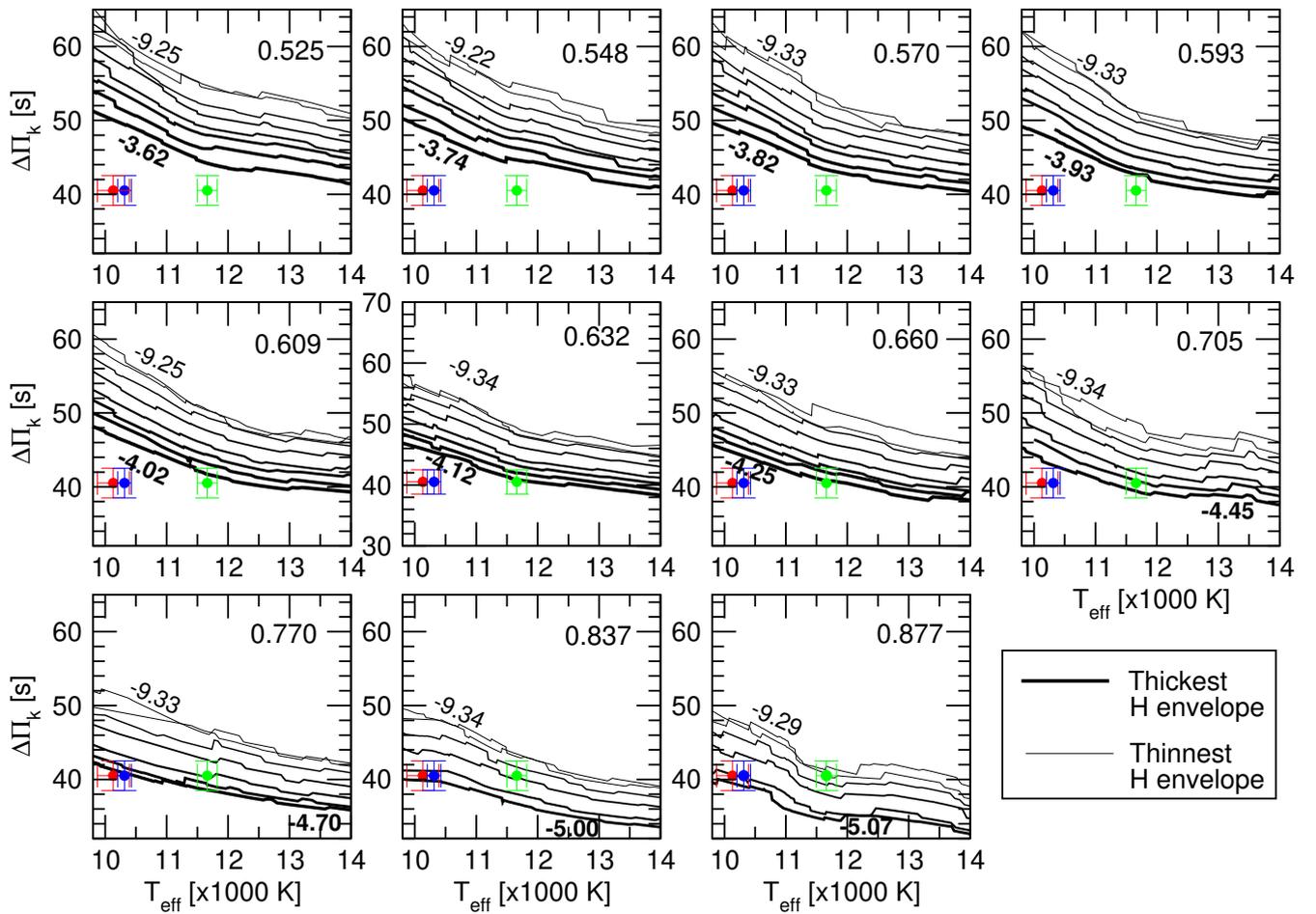}
    \caption{Average of the computed dipole ($\ell=1$) period spacings, $\overline{\Delta \Pi_k}$, in terms of the effective temperature, for different stellar masses in solar units (numbers at the top right corner of each panel) and thicknesses of the H envelope \cite[see Table 4 of][for the specific values of $\log(M_{\rm H}/M_{\star})$]{2023MNRAS.526.2846U}  drawn with different colours. In each panel, we include numbers along two curves, which correspond to the value of $\log(M_{\rm H}/M_{\star})$ for the thickest and the thinnest H envelopes for each stellar mass value. The location of \wda\ is emphasized with an circle with error bars with effective temperatures $T_{\rm eff}= 10\,131\pm 260$\,K \citep[red symbol;][]{2018ApJS..239...26L}, $T_{\rm eff}= 10\,313\pm 111$\,K \citep[blue symbol;][]{2024MNRAS.527.8687O}, and $T_{\rm eff}= 11\,660\pm 163$\,K (green symbol; Section~\ref{Joint_fit} of this work), and a period spacing $\Delta \Pi= 40.51\pm2.00$~s (Section \ref{ps_tests}).}
\label{fig:acps}
\end{figure*}

A practical approach to estimate the stellar mass of pulsating WD stars involves comparing the observed period spacing ($\Delta \Pi$) with the average of the calculated period spacings ($\overline{\Delta \Pi_{k}}$) \citep{2019A&ARv..27....7C}. The average is determined using the formula $\overline{\Delta \Pi_{k}}= (n-1)^{-1} \sum_k \Delta \Pi_{k}$, where the "forward" period spacing ($\Delta \Pi_{k}$) is defined as $\Delta \Pi_{k}= \Pi_{k+1}-\Pi_{k}$ (with $k$ representing the radial order) and $n$ is the number of periods computed that fall within the range of observed periods.  It is important to note that this method for determining stellar mass depends on the spectroscopic effective temperature, and the results are inevitably influenced by the uncertainties associated with $T_{\rm eff}$. The method mentioned leverages the fact that, in general, the period spacing of pulsating WD stars is mainly influenced by stellar mass and effective temperature, with only a minor dependence on the thickness of the He envelope for DBV stars or the O/C/He envelope for GW Vir stars \citep[see, e.g.,][]{1990ApJS...72..335T}. However, this technique cannot be directly applied to DAV stars for mass estimation, as the period spacing in these stars is influenced by $M_{\star}$, $T_{\rm eff}$, and the mass of the H envelope $M_{\rm H}$, with similar sensitivity, leading to multiple combinations of these three parameters that yield the same period spacing. Therefore, we can only provide a possible range of stellar masses for \wda\ based on the period spacing.

We calculated the mean of the period spacings for $\ell=1$, denoted as $\overline{\Delta \Pi_{k}}$, employing the {\tt LP-PUL} pulsation code \citep{2006A&A...454..863C}, in terms  of the effective temperature across all the stellar masses considered and the thicknesses of the H envelope \cite[see Table 4 of][]{2023MNRAS.526.2846U}. The analysed period range was established between 300 and 1600 seconds, encompassing the typical periods observed in the target star \wda. The effective temperature of \wda\ according to this work (see Sect.~\ref{Joint_fit}) is $T_{\rm eff}= 11\,660\pm163$\,K on average. However, there are other measurements of $T_{\rm eff}$ for this DAV star in the literature, such as $T_{\rm eff}= 10\,131\pm 260$\,K \citep{2018ApJS..239...26L}, $T_{\rm eff}= 10\,313\pm 111$\,K \citep{2024MNRAS.527.8687O}, $T_{\rm eff}= 11\,617\pm 70$\,K \citep{2018MNRAS.473.3693G}, and $T_{\rm eff}= 11\,600\pm 200$\,K \citep{Munday2024}. In particular, the value of $T_{\rm eff}$ derived in the present study is almost the same as the values derived by \cite{2018MNRAS.473.3693G} and \cite{Munday2024}. 

As there is no definitive effective temperature measurement available for this star, we used multiple estimates to cover the possible range of $T_{\rm eff}$. The results are illustrated in Fig.~\ref{fig:acps}, showing $\overline{\Delta \Pi_{k}}$ for various stellar masses (specified at the top right corner of each panel), represented by curves with different thicknesses corresponding to the diverse values of $M_{\rm H}$. To enhance clarity, we only labelled the extreme H-envelope thickness values for each stellar mass, using thick and thin black curves. The position of \wda, marked by a small circle with error bars, was examined with three spectroscopic effective temperature values: $T_{\rm eff}= 10\,131\pm 260$\,K \citep{2018ApJS..239...26L}, $T_{\rm eff}= 10\,313\pm 111$\,K \citep{2024MNRAS.527.8687O}, representative of most of the $T_{\rm eff}$ determinations, which point to low effective temperatures, and $T_{\rm eff}= 11\,660\pm 163$\,K (this paper), along with a period spacing of $\Delta \Pi= 40.51\pm 0.14$\,s.

Upon analysis of the plot, we infer that, based on the period spacing and $T_{\rm eff}$, the stellar mass of \wda\ likely falls between $0.570\,M_{\odot}$ (with a thick H envelope of $\log(M_{\rm H}/M_{\star})= -3.82$) and $0.877\,M_{\odot}$ (with a very thin H envelope of $\log(M_{\rm H}/M_{\star})= -9.29$) if the effective temperature is high (green dot in Fig. \ref{fig:acps}). In contrast, if the effective temperature of \wda\ is lower (as indicated by the red and blue dots in Fig. \ref{fig:acps}), then the stellar mass would probably exceed $0.705 M_{\sun}$ with a thick envelope of H ($\log(M_{\rm H}/M_{\star})= -4.45$).  In summary, based on the period spacing and effective temperature, the stellar mass of \wda\ would be larger than $\sim 0.57 M_{\sun}$.

\subsection{Asteroseismic period-to-period fits}
\label{apf}

In this section, we try 
to find an evolutionary model that best matches the theoretical periods with the individual pulsation periods detected for WD~1310+583. The quality of the fit is assessed by evaluating the quality function defined as follows:

\begin{equation}
\sigma ^2(M_{\star},M_H,T_{\rm eff})=\frac{1}{N} \sum_{i=1}^{N} \rm min[(\Pi_i^{O}-\Pi_k^{th})^2], 
\label{eq:quality}
\end{equation}

Here, $N$ represents the number of detected modes, $\Pi_i^{\rm O}$ are the observed periods, and $\Pi_k^{\rm th}$ are the theoretically computed periods (where $k$ is the radial order). The best-fitting model, if it exists, is chosen by identifying the minimum value of $\sigma^2$.

\begin{table*}[]
  \centering
\caption{Parameters of the best-fit models.}
\begin{tabular}{cccccccc}
\hline
\hline
Model & $M_{\star}$ & $\log g$ & $T_{\rm eff}$  & $\log\left(\frac{M_{\rm H}}{M_{\star}}\right)$  &  $\sigma^2 (s^2)$ & $\ell= 1$ & $d$\\
\#      &   ($M_{\odot}$)   &  ($\rm cm~ s^{-2}$)     &  (K)   &       &     &   & (pc)  \\
\hline
1&0.609   &8.03   &11\,441   &-4.02   &26.52 &21 &27.65   \\
2&0.837   & 8.39  &11\,147   &-5.35   &27.07  & 19& 20.74  \\
3&0.632   &8.06   &11\,702   &-5.35   &26.03  & 22&27.75   \\
\hline
\end{tabular}
\label{tab:best-fits}
\end{table*}

We use the same grid of CO-core WD models as in \cite{2023MNRAS.526.2846U}, which contemplate evolutionary sequences of stellar masses in the range $0.525\leq M_{\star}/M_{\odot} \leq 0.877$, with effective temperature $10\,000\,{\rm K} \leq T_{\rm eff}\leq 13\,000$\,K, and varying the total hydrogen content $-9\lesssim \log(M_{\rm H}/M_{\star}) \lesssim -4$. 

For the period-to-period fit, we examined the frequency spectrum of WD~1310+583 and used the confirmed triplets and those consecutive overtones derived as $\ell= 1$ as input priors; see the Appendix~\ref{app:A} and Table~\ref{tabl:rot}. We considered the central components (with $m=0$), specifically the 8 observed periods (793.4, 837.6, 873.8, 919.3, 953.6, 997.1, 1038.4 and 1078.4 s), which were assigned $\ell=1$ for the analysis.

\begin{figure}
	\includegraphics[width=0.5\textwidth]{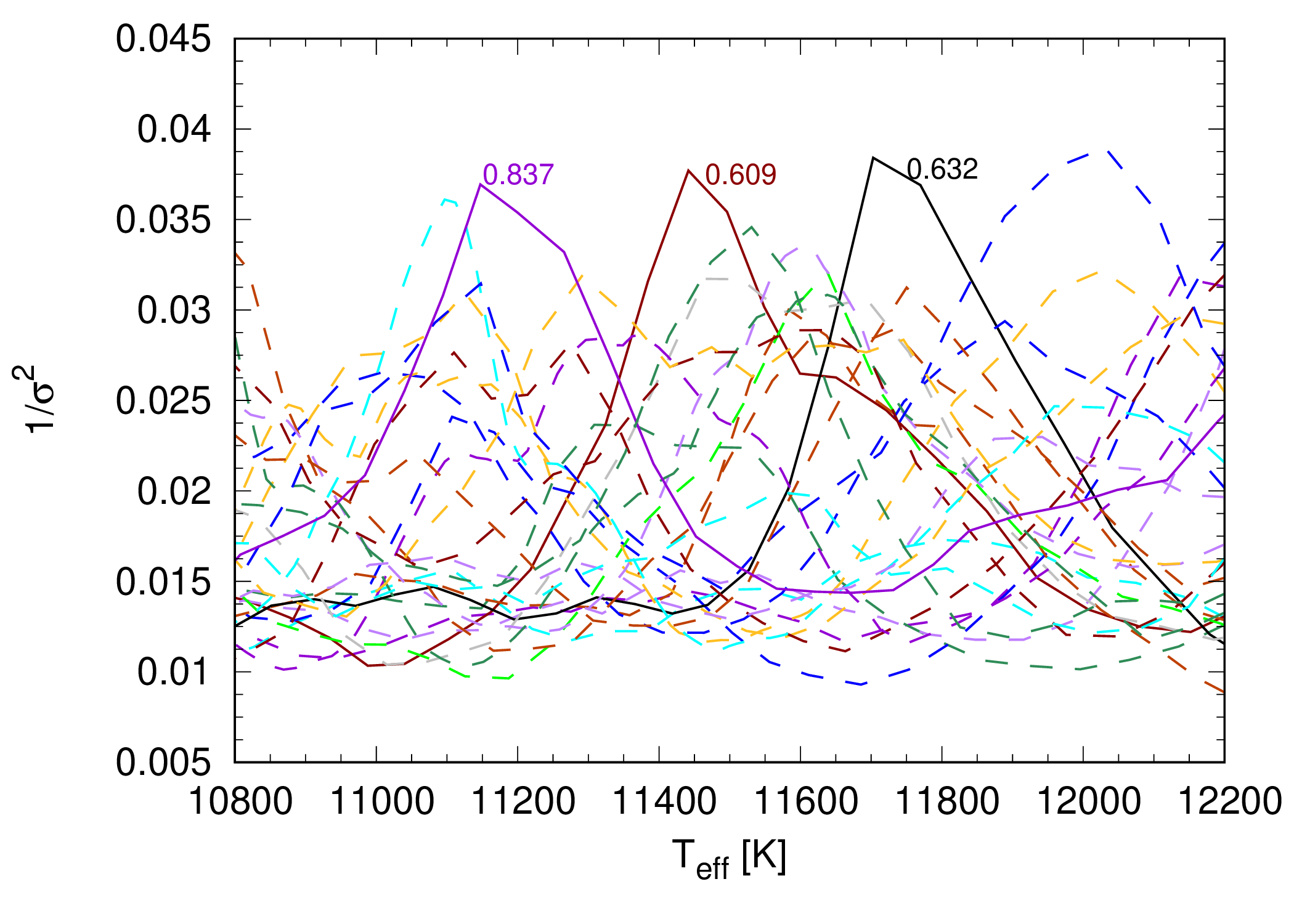}
    \caption{Inverse of the squared quality as a function $\sigma$ defined in Eq.~\ref{eq:quality} in terms 
of the effective temperature. Solid (dashed) lines represent the value for the selected (discarded) models (see Table \ref{tab:best-fits}).}
\label{fig:quality}
\end{figure}

In Fig.~\ref{fig:quality} we show the value of the inverse of the squared quality function for our models, in the range of effective temperatures of interest. 
With solid lines, we highlight the quality function for the models that best match the observed periods, as indicated by the maximum in the curve. The two models represented by the maxima of the dashed blue (cyan) 
curves next to the solid 
curves were discarded because of their hot (cold) effective temperatures.
The three best-fit models highlighted represent only marginally better fits than other models, something that prevents us from considering them as seismic measurements of the properties of \wda. Indeed, the three individual highest peaks are not statistically compelling enough for their parameters to be worth discussing as properties of this star. In view of this,
from now on we will limit ourselves to describing the properties of tentative seismological solutions for \wda.

The most important stellar parameters of our tentative asteroseismic models are shown in Table~\ref{tab:best-fits}. 
Given the $T_{\rm eff}$ values obtained in this work, which considers the binary nature of this system, we focus our 
analysis in the vicinity of $T_{\rm eff}\sim 11\, 600$\,K. Based on 
our results,
it is more likely that WD~1310+583 would have 
a thick H-envelope.
 However, the more massive solution exhibits a significant discrepancy with our spectroscopic determination of $\log g$ and the Gaia distance; therefore, we reject this model as a possible solution.

Always bearing in mind that none of these solutions is formally statistically significant, we can consider model~3 as a possible description of the structure of \wda\ since its mass aligns with the results of the mean period spacing analysis and provides the best agreement between the theoretical and observed periods.
Our tentative asteroseismological model is characterised by total H- and He-content of  $4.46\times 10^{-6} M_{*}$  and $1.75\times 10^{-2} M_{*}$, respectively, and a luminosity of $\rm L/L_{\odot}=0.25\times 10^{-2}$.
We derived the internal uncertainties in the most important stellar parameters from the model grid resolution $\chi_{M_{\star}}=0.028~M_{\odot}$, $\chi_{T_{\rm eff}}= 70$\,K, $\chi_{\log g}=0.05$, $\chi_{L_{\star}/L_{\odot}}=2\times 10^{-4}$ and $\chi_{R_{\star}/R_{\odot}}=1.2\times 10^{-4}$.

Based on the stellar parameters derived from our tentative representative model, we estimate the asteroseismic distance of \wda. 
From the derived effective temperature and the logarithm of surface gravity, we calculated the absolute magnitude in the Gaia $G$ band (D. Koester, personal communication). For the 
model with a mass of 0.632\,$M_{\odot}$, we find that the absolute magnitude is $M_{\rm G} = 11.85$ mag. From the apparent magnitude obtained by the Gaia Data Release 3 (DR3) Archive\footnote{\url{https://gea.esac.esa.int/archive/}} for WD 1310+583 ($m_{\rm G} =14.07$ mag),  we obtain an asteroseismic distance of $d = 27.75^{+0.17}_{-0.15}$\,pc.
We found that our distance is appreciably smaller compared to the Gaia distance \citep{2021yCat.1352....0B}, which reports of $d = 30.79\pm 0.2$\,pc.

\section{Discussion}

WD~1310+583 was one of the targets observed from Konkoly Observatory as part of a survey aimed at discovering new pulsating white dwarf stars potentially detectable by TESS (see \citealt{2018MNRAS.478.2676B}). That study presented observations obtained over eight nights, complemented by a preliminary asteroseismic analysis of the star. They identified 17 significant frequencies, seven of which were found to be independent pulsation modes, as also summarised in the introduction section of this paper. Preliminary asteroseismic modelling was performed for six of these modes using the White Dwarf Evolution Code (\textsc{WDEC}; see \citealt{2008ApJ...675.1505B}). Four of the six modes were also detected in the TESS data, within the limits of the observational uncertainties.

The preliminary asteroseismic analysis presented by \citet{2018MNRAS.478.2676B} aimed to constrain the stellar mass, the effective temperature, and the mass of the hydrogen layer. Adapting the observed periods for six modes and assuming that at least four of them, including the dominant one, correspond to dipole modes ($\ell = 1$), the following parameters were derived: $M_* = 0.74\,M_\odot$, $T_{\mathrm{eff}} = 11\,600\,\mathrm{K}$, and $-\log M_{\mathrm{H}} = 4.0$. This mass is approximately $0.1\,M_\odot$ higher than that obtained from fully evolutionary models (this work), while the hydrogen layer mass is roughly two orders of magnitude larger. However, the effective temperatures derived from the two modelling approaches are quite similar (11\,600~K vs. 11\,702\,K).

\citet{2018MNRAS.478.2676B} also reported possible values of the rotational frequency splittings and the corresponding rotation periods. The stellar rotation period could be either 5~h or 1.3~d. It should be noted that the latter value is very close to the 1.18~d obtained from the TESS measurements.

The comparison between our asteroseismic results and previous analyses highlights the typical challenges in reconciling spectroscopic and seismological determinations of white dwarf parameters, particularly for DAVs (see \citealt{2024A&A...691A.194C}). In this case, the asteroseismic mass derived from the WDEC modelling exceeds the spectroscopic mass by about $0.1\,M_\odot$, a discrepancy that falls within the range commonly observed in pulsating DAs. Such differences may arise from systematic uncertainties that affect both methods. Spectroscopic parameters ($T_{\mathrm{eff}}$, $\log g$) depend on the atmosphere models adopted and the treatment of line broadening and convection, which significantly influence the Balmer line profiles (\citealt{2022PhR...988....1S}). As shown by \citet{2017PhDT........20F}, unaccounted for systematics such as imperfect flux calibration, extinction corrections, or low signal-to-noise ratios can introduce biases in $T_{\mathrm{eff}}$ and $\log g$, thereby affecting the derived masses. On the other hand, asteroseismic masses depend sensitively on mode identification and on the adopted model grids and internal chemical stratification profiles \citep{2018A&A...613A..46D,2017A&A...599A..21D}. For WD~1310+583, the assumption that most detected modes correspond to $\ell=1$ may have influenced the 
stellar structure parameters of our tentative asteroseismological model, since even a small number of $\ell=2$ modes can alter the inferred mass and thickness of the hydrogen layer.

We also emphasise that, as Fig.~\ref{fig:quality} clearly shows, our selected model is only a tentative solution. The best-fit models are only marginally better fits than the other models, which means that the models presented in Table~\ref{tab:best-fits} are not statistically significant, so we cannot find a robust seismological solution. Although the star shows a rich pulsation spectrum with robustly identified modes, this does not mean that we can find a representative model, highlighting the need for improvements on the modelling side and the asteroseismological method employed, as also discussed in \citet{2024A&A...691A.194C}. 

Our analysis of WD~1310+583 exemplifies the broader trend observed among DAVs, where the spectroscopic mass may overestimate or underestimate the seismological mass depending on the interplay of uncertainties in both approaches. The degeneracy between the core and envelope structure in high-overtone $g$-modes (see \citealt{2003MNRAS.344..657M,2017A&A...598A.109G}) also contributes to potential ambiguities in the derived stellar parameters, even when several independent pulsation periods are available. In particular, the number of detected modes is not necessarily indicative of the reliability of the seismological solution; rather, it is their distribution in radial order and the degree of mode trapping that determine the diagnostic power of the dataset.

\section{Summary}
\label{sect:sum}

This study presents an asteroseismic investigation of the ZZ~Ceti star \wda. A prerequisite for asteroseismic modelling is the identification of the star’s normal pulsation modes. These modes were derived from 120-second cadence photometric data obtained with the Transiting Exoplanet Survey Satellite (TESS), supplemented by relevant parameters available in the literature.

Additional constraints were provided by spectroscopic analysis, based on data collected with the Cosmic Origins Spectrograph onboard the Hubble Space Telescope. We also examined the possible presence of rotationally split multiplets.

By means of an asteroseismic analysis, we derived a
 suggestive asteroseismological model characterised by the following physical parameters: $M_*=0.632\,M_{\odot}$, $T_{\mathrm{eff}}=11\,702\,$K, the stellar mass being compatible with the predictions of the period spacing ($M_{\star} \gtrsim 0.6\,M_{\sun}$). The corresponding asteroseismic distance (27.75\,pc) is smaller compared to the geometric distance inferred from Gaia astrometry (30.79\,pc). 

Continued operation of the TESS mission will enable not only similar in-depth case studies but also ensemble analyses of white dwarf pulsators and the discovery of new compact variables.

\begin{acknowledgements}

The authors thank the anonymous referee for constructive comments and recommendations on the manuscript.

The authors acknowledge Michael H. Montgomery (Department of Astronomy, University of Texas at Austin; McDonald Observatory, Fort Davis) for the useful early discussions. The authors also acknowledge Mukremin Kilic (Homer L. Dodge Department of Physics and Astronomy, University of Oklahoma), Antoine Bedard (Department of Physics, University of Warwick) and Detlev Koester (Institut f\"ur Theoretische Physik und Astrophysik, Universit\"at Kiel) for their helpful discussion. Zs.B. and \'A.S. acknowledge the financial support of the KKP-137523 `SeismoLab' \'Elvonal grant of the Hungarian Research, Development and Innovation Office (NKFIH). M.U. gratefully acknowledges support from the Research Foundation Flanders (FWO) through a Junior Postdoctoral Fellowship (grant agreement No: 1247624N).

This paper includes data collected with the TESS mission, obtained from the MAST data archive at the Space Telescope Science Institute (STScI). Funding for the TESS mission is provided by the NASA Explorer Programme. STScI is operated by the Association of Universities for Research in Astronomy, Inc., under NASA contract NAS 5–26555.

\end{acknowledgements}



\bibliographystyle{aa} 
\bibliography{aa56826-25} 

\begin{thebibliography}{53}
\expandafter\ifx\csname natexlab\endcsname\relax\def\natexlab#1{#1}\fi

\bibitem[{{Ahumada} {et~al.}(2020){Ahumada}, {Allende Prieto}, {Almeida}, {Anders}, {Anderson}, {Andrews}, {Anguiano}, {Arcodia}, {Armengaud}, {Aubert}, {Avila}, {Avila-Reese}, {Badenes}, {Balland}, {Barger}, {Barrera-Ballesteros}, {Basu}, {Bautista}, {Beaton}, {Beers}, {Benavides}, {Bender}, {Bernardi}, {Bershady}, {Beutler}, {Bidin}, {Bird}, {Bizyaev}, {Blanc}, {Blanton}, {Boquien}, {Borissova}, {Bovy}, {Brandt}, {Brinkmann}, {Brownstein}, {Bundy}, {Bureau}, {Burgasser}, {Burtin}, {Cano-D{\'\i}az}, {Capasso}, {Cappellari}, {Carrera}, {Chabanier}, {Chaplin}, {Chapman}, {Cherinka}, {Chiappini}, {Doohyun Choi}, {Chojnowski}, {Chung}, {Clerc}, {Coffey}, {Comerford}, {Comparat}, {da Costa}, {Cousinou}, {Covey}, {Crane}, {Cunha}, {Ilha}, {Dai}, {Damsted}, {Darling}, {Davidson}, {Davies}, {Dawson}, {De}, {de la Macorra}, {De Lee}, {Queiroz}, {Deconto Machado}, {de la Torre}, {Dell'Agli}, {du Mas des Bourboux}, {Diamond-Stanic}, {Dillon}, {Donor}, {Drory}, {Duckworth}, {Dwelly}, {Ebelke}, {Eftekharzadeh}, {Davis
  Eigenbrot}, {Elsworth}, {Eracleous}, {Erfanianfar}, {Escoffier}, {Fan}, {Farr}, {Fern{\'a}ndez-Trincado}, {Feuillet}, {Finoguenov}, {Fofie}, {Fraser-McKelvie}, {Frinchaboy}, {Fromenteau}, {Fu}, {Galbany}, {Garcia}, {Garc{\'\i}a-Hern{\'a}ndez}, {Garma Oehmichen}, {Ge}, {Geimba Maia}, {Geisler}, {Gelfand}, {Goddy}, {Gonzalez-Perez}, {Grabowski}, {Green}, {Grier}, {Guo}, {Guy}, {Harding}, {Hasselquist}, {Hawken}, {Hayes}, {Hearty}, {Hekker}, {Hogg}, {Holtzman}, {Horta}, {Hou}, {Hsieh}, {Huber}, {Hunt}, {Ider Chitham}, {Imig}, {Jaber}, {Jimenez Angel}, {Johnson}, {Jones}, {J{\"o}nsson}, {Jullo}, {Kim}, {Kinemuchi}, {Kirkpatrick}, {Kite}, {Klaene}, {Kneib}, {Kollmeier}, {Kong}, {Kounkel}, {Krishnarao}, {Lacerna}, {Lan}, {Lane}, {Law}, {Le Goff}, {Leung}, {Lewis}, {Li}, {Lian}, {Lin}, {Long}, {Longa-Pe{\~n}a}, {Lundgren}, {Lyke}, {Mackereth}, {MacLeod}, {Majewski}, {Manchado}, {Maraston}, {Martini}, {Masseron}, {Masters}, {Mathur}, {McDermid}, {Merloni}, {Merrifield}, {M{\'e}sz{\'a}ros}, {Miglio}, {Minniti},
  {Minsley}, {Miyaji}, {Mohammad}, {Mosser}, {Mueller}, {Muna}, {Mu{\~n}oz-Guti{\'e}rrez}, {Myers}, {Nadathur}, {Nair}, {Nandra}, {Correa do Nascimento}, {Nevin}, {Newman}, {Nidever}, {Nitschelm}, {Noterdaeme}, {O'Connell}, {Olmstead}, {Oravetz}, {Oravetz}, {Osorio}, {Pace}, {Padilla}, {Palanque-Delabrouille}, {Palicio}, {Pan}, {Pan}, {Parker}, {Paviot}, {Peirani}, {Ram{\'r}ez}, {Penny}, {Percival}, {Perez-Fournon}, {P{\'e}rez-R{\`a}fols}, {Petitjean}, {Pieri}, {Pinsonneault}, {Poovelil}, {Povick}, {Prakash}, {Price-Whelan}, {Raddick}, {Raichoor}, {Ray}, {Rembold}, {Rezaie}, {Riffel}, {Riffel}, {Rix}, {Robin}, {Roman-Lopes}, {Rom{\'a}n-Z{\'u}{\~n}iga}, {Rose}, {Ross}, {Rossi}, {Rowlands}, {Rubin}, {Salvato}, {S{\'a}nchez}, {S{\'a}nchez-Menguiano}, {S{\'a}nchez-Gallego}, {Sayres}, {Schaefer}, {Schiavon}, {Schimoia}, {Schlafly}, {Schlegel}, {Schneider}, {Schultheis}, {Schwope}, {Seo}, {Serenelli}, {Shafieloo}, {Shamsi}, {Shao}, {Shen}, {Shetrone}, {Shirley}, {Silva Aguirre}, {Simon}, {Skrutskie}, {Slosar},
  {Smethurst}, {Sobeck}, {Sodi}, {Souto}, {Stark}, {Stassun}, {Steinmetz}, {Stello}, {Stermer}, {Storchi-Bergmann}, {Streblyanska}, {Stringfellow}, {Stutz}, {Su{\'a}rez}, {Sun}, {Taghizadeh-Popp}, {Talbot}, {Tayar}, {Thakar}, {Theriault}, {Thomas}, {Thomas}, {Tinker}, {Tojeiro}, {Toledo}, {Tremonti}, {Troup}, {Tuttle}, {Unda-Sanzana}, {Valentini}, {Vargas-Gonz{\'a}lez}, {Vargas-Maga{\~n}a}, {V{\'a}zquez-Mata}, {Vivek}, {Wake}, {Wang}, {Weaver}, {Weijmans}, {Wild}, {Wilson}, {Wilson}, {Wolthuis}, {Wood-Vasey}, {Yan}, {Yang}, {Y{\`e}che}, {Zamora}, {Zarrouk}, {Zasowski}, {Zhang}, {Zhao}, {Zhao}, {Zheng}, {Zheng}, {Zhu}, \& {Zou}}]{SDSSdr16}
{Ahumada}, R., {Allende Prieto}, C., {Almeida}, A., {et~al.} 2020, \apjs, 249, 3

\bibitem[{{Althaus} {et~al.}(2010){Althaus}, {C{\'o}rsico}, {Isern}, \& {Garc{\'{\i}}a-Berro}}]{2010A&ARv..18..471A}
{Althaus}, L.~G., {C{\'o}rsico}, A.~H., {Isern}, J., \& {Garc{\'{\i}}a-Berro}, E. 2010, \aapr, 18, 471

\bibitem[{{Bailer-Jones} {et~al.}(2021){Bailer-Jones}, {Rybizki}, {Fouesneau}, {Demleitner}, \& {Andrae}}]{2021yCat.1352....0B}
{Bailer-Jones}, C.~A.~L., {Rybizki}, J., {Fouesneau}, M., {Demleitner}, M., \& {Andrae}, R. 2021, {VizieR Online Data Catalog: Distances to 1.47 billion stars in Gaia EDR3 (Bailer-Jones+, 2021)}, VizieR On-line Data Catalog: I/352. Originally published in: 2021AJ....161..147B

\bibitem[{{B{\'e}dard} {et~al.}(2020){B{\'e}dard}, {Bergeron}, {Brassard}, \& {Fontaine}}]{Bedard2020}
{B{\'e}dard}, A., {Bergeron}, P., {Brassard}, P., \& {Fontaine}, G. 2020, \apj, 901, 93

\bibitem[{{Bell} {et~al.}(2019){Bell}, {C{\'o}rsico}, {Bischoff-Kim}, {Althaus}, {Bradley}, {Calcaferro}, {Montgomery}, {Uzundag}, {Baran}, {Bogn{\'a}r}, {Charpinet}, {Ghasemi}, \& {Hermes}}]{2019A&A...632A..42B}
{Bell}, K.~J., {C{\'o}rsico}, A.~H., {Bischoff-Kim}, A., {et~al.} 2019, \aap, 632, A42

\bibitem[{{Bell} {et~al.}(2015){Bell}, {Hermes}, {Bischoff-Kim}, {Moorhead}, {Montgomery}, {{\O}stensen}, {Castanheira}, \& {Winget}}]{2015ApJ...809...14B}
{Bell}, K.~J., {Hermes}, J.~J., {Bischoff-Kim}, A., {et~al.} 2015, \apj, 809, 14

\bibitem[{{Bell} {et~al.}(2016){Bell}, {Hermes}, {Montgomery}, {Gentile Fusillo}, {Raddi}, {G{\"a}nsicke}, {Winget}, {Dennihy}, {Gianninas}, {Tremblay}, {Chote}, \& {Winget}}]{2016ApJ...829...82B}
{Bell}, K.~J., {Hermes}, J.~J., {Montgomery}, M.~H., {et~al.} 2016, \apj, 829, 82

\bibitem[{{Bell} {et~al.}(2017){Bell}, {Hermes}, {Montgomery}, {Winget}, {Gentile Fusillo}, {Raddi}, \& {G{\"a}nsicke}}]{2017ASPC..509..303B}
{Bell}, K.~J., {Hermes}, J.~J., {Montgomery}, M.~H., {et~al.} 2017, in Astronomical Society of the Pacific Conference Series, Vol. 509, 20th European White Dwarf Workshop, ed. P.-E. {Tremblay}, B.~{Gaensicke}, \& T.~{Marsh}, 303

\bibitem[{{Bischoff-Kim} {et~al.}(2008){Bischoff-Kim}, {Montgomery}, \& {Winget}}]{2008ApJ...675.1505B}
{Bischoff-Kim}, A., {Montgomery}, M.~H., \& {Winget}, D.~E. 2008, \apj, 675, 1505

\bibitem[{{Bogn{\'a}r} {et~al.}(2018){Bogn{\'a}r}, {Kalup}, {S{\'o}dor}, {Charpinet}, \& {Hermes}}]{2018MNRAS.478.2676B}
{Bogn{\'a}r}, Z., {Kalup}, C., {S{\'o}dor}, {\'A}., {Charpinet}, S., \& {Hermes}, J.~J. 2018, \mnras, 478, 2676

\bibitem[{{Bogn{\'a}r} {et~al.}(2020){Bogn{\'a}r}, {Kawaler}, {Bell}, {Schrandt}, {Baran}, {Bradley}, {Hermes}, {Charpinet}, {Handler}, {Mullally}, {Murphy}, {Raddi}, {S{\'o}dor}, {Tremblay}, {Uzundag}, \& {Zong}}]{2020A&A...638A..82B}
{Bogn{\'a}r}, Z., {Kawaler}, S.~D., {Bell}, K.~J., {et~al.} 2020, \aap, 638, A82

\bibitem[{{Bogn{\'a}r} {et~al.}(2023){Bogn{\'a}r}, {S{\'o}dor}, {Clark}, \& {Kawaler}}]{2023A&A...674A.204B}
{Bogn{\'a}r}, Z., {S{\'o}dor}, {\'A}., {Clark}, I.~R., \& {Kawaler}, S.~D. 2023, \aap, 674, A204

\bibitem[{{Brickhill}(1991)}]{1991MNRAS.251..673B}
{Brickhill}, A.~J. 1991, \mnras, 251, 673

\bibitem[{{Calcaferro} {et~al.}(2024){Calcaferro}, {C{\'o}rsico}, {Uzundag}, {Althaus}, {Kepler}, \& {Werner}}]{2024A&A...691A.194C}
{Calcaferro}, L.~M., {C{\'o}rsico}, A.~H., {Uzundag}, M., {et~al.} 2024, \aap, 691, A194

\bibitem[{{Castro-Ginard} {et~al.}(2024){Castro-Ginard}, {Penoyre}, {Casey}, {Brown}, {Belokurov}, {Cantat-Gaudin}, {Drimmel}, {Fouesneau}, {Khanna}, {Kurbatov}, {Price-Whelan}, {Rix}, \& {Smart}}]{2024A&A...688A...1C}
{Castro-Ginard}, A., {Penoyre}, Z., {Casey}, A.~R., {et~al.} 2024, \aap, 688, A1

\bibitem[{{Chambers} \& {Pan-STARRS Team}(2018)}]{Panstarrs}
{Chambers}, K. \& {Pan-STARRS Team}. 2018, in American Astronomical Society Meeting Abstracts, Vol. 231, American Astronomical Society Meeting Abstracts \#231, 102.01

\bibitem[{{C{\'o}rsico}(2020)}]{2020FrASS...7...47C}
{C{\'o}rsico}, A.~H. 2020, Frontiers in Astronomy and Space Sciences, 7, 47

\bibitem[{{C{\'o}rsico} \& {Althaus}(2006)}]{2006A&A...454..863C}
{C{\'o}rsico}, A.~H. \& {Althaus}, L.~G. 2006, \aap, 454, 863

\bibitem[{{C{\'o}rsico} {et~al.}(2019){C{\'o}rsico}, {Althaus}, {Miller Bertolami}, \& {Kepler}}]{2019A&ARv..27....7C}
{C{\'o}rsico}, A.~H., {Althaus}, L.~G., {Miller Bertolami}, M.~M., \& {Kepler}, S.~O. 2019, \aapr, 27, 7

\bibitem[{{Cukanovaite} {et~al.}(2021){Cukanovaite}, {Tremblay}, {Bergeron}, {Freytag}, {Ludwig}, \& {Steffen}}]{ElenaCukanovaite2021_3D_DB}
{Cukanovaite}, E., {Tremblay}, P.-E., {Bergeron}, P., {et~al.} 2021, \mnras, 501, 5274

\bibitem[{{De Ger{\'o}nimo} {et~al.}(2017){De Ger{\'o}nimo}, {Althaus}, {C{\'o}rsico}, {Romero}, \& {Kepler}}]{2017A&A...599A..21D}
{De Ger{\'o}nimo}, F.~C., {Althaus}, L.~G., {C{\'o}rsico}, A.~H., {Romero}, A.~D., \& {Kepler}, S.~O. 2017, \aap, 599, A21

\bibitem[{{De Ger{\'o}nimo} {et~al.}(2018){De Ger{\'o}nimo}, {Althaus}, {C{\'o}rsico}, {Romero}, \& {Kepler}}]{2018A&A...613A..46D}
{De Ger{\'o}nimo}, F.~C., {Althaus}, L.~G., {C{\'o}rsico}, A.~H., {Romero}, A.~D., \& {Kepler}, S.~O. 2018, \aap, 613, A46

\bibitem[{{Dolez} \& {Vauclair}(1981)}]{1981A&A...102..375D}
{Dolez}, N. \& {Vauclair}, G. 1981, \aap, 102, 375

\bibitem[{{Fontaine} \& {Brassard}(2008)}]{2008PASP..120.1043F}
{Fontaine}, G. \& {Brassard}, P. 2008, \pasp, 120, 1043

\bibitem[{{Fuchs}(2017)}]{2017PhDT........20F}
{Fuchs}, J.~T. 2017, PhD thesis, University of North Carolina, Chapel Hill

\bibitem[{{Gaia Collaboration}(2022)}]{2022yCat.1355....0G}
{Gaia Collaboration}. 2022, {VizieR Online Data Catalog: Gaia DR3 Part 1. Main source (Gaia Collaboration, 2022)}, VizieR On-line Data Catalog: I/355. Originally published in: doi:10.1051/0004-63

\bibitem[{{Gentile Fusillo} {et~al.}(2021){Gentile Fusillo}, {Tremblay}, {Cukanovaite}, {Vorontseva}, {Lallement}, {Hollands}, {G{\"a}nsicke}, {Burdge}, {McCleery}, \& {Jordan}}]{2021MNRAS.508.3877G}
{Gentile Fusillo}, N.~P., {Tremblay}, P.~E., {Cukanovaite}, E., {et~al.} 2021, \mnras, 508, 3877

\bibitem[{{Gentile Fusillo} {et~al.}(2018){Gentile Fusillo}, {Tremblay}, {Jordan}, {G{\"a}nsicke}, {Kalirai}, \& {Cummings}}]{2018MNRAS.473.3693G}
{Gentile Fusillo}, N.~P., {Tremblay}, P.~E., {Jordan}, S., {et~al.} 2018, \mnras, 473, 3693

\bibitem[{{Giammichele} {et~al.}(2017){Giammichele}, {Charpinet}, {Brassard}, \& {Fontaine}}]{2017A&A...598A.109G}
{Giammichele}, N., {Charpinet}, S., {Brassard}, P., \& {Fontaine}, G. 2017, \aap, 598, A109

\bibitem[{{Goldreich} \& {Wu}(1999)}]{1999ApJ...511..904G}
{Goldreich}, P. \& {Wu}, Y. 1999, \apj, 511, 904

\bibitem[{{Handler} {et~al.}(1997){Handler}, {Pikall}, {O'Donoghue}, {Buckley}, {Vauclair}, {Chevreton}, {Giovannini}, {Kepler}, {Goode}, {Provencal}, {Wood}, {Clemens}, {O'Brien}, {Nather}, {Winget}, {Kleinman}, {Kanaan}, {Watson}, {Nitta}, {Montgomery}, {Klumpe}, {Bradley}, {Sullivan}, {Wu}, {Marar}, {Seetha}, {Ashoka}, {Mahra}, {Bhat}, {Babu}, {Leibowitz}, {Hemar}, {Ibbetson}, {Mashal}, {Meistas}, {Dziembowski}, {Pamyatnykh}, {Moskalik}, {Zola}, {Pajdosz}, {Krzesinski}, {Solheim}, {Bard}, {Massacand}, {Breger}, {Gelbmann}, {Paunzen}, \& {North}}]{1997MNRAS.286..303H}
{Handler}, G., {Pikall}, H., {O'Donoghue}, D., {et~al.} 1997, \mnras, 286, 303

\bibitem[{{Hermes} {et~al.}(2017){Hermes}, {G{\"a}nsicke}, {Kawaler}, {Greiss}, {Tremblay}, {Gentile Fusillo}, {Raddi}, {Fanale}, {Bell}, {Dennihy}, {Fuchs}, {Dunlap}, {Clemens}, {Montgomery}, {Winget}, {Chote}, {Marsh}, \& {Redfield}}]{2017ApJS..232...23H}
{Hermes}, J.~J., {G{\"a}nsicke}, B.~T., {Kawaler}, S.~D., {et~al.} 2017, \apjs, 232, 23

\bibitem[{{Hermes} {et~al.}(2015){Hermes}, {Montgomery}, {Bell}, {Chote}, {G{\"a}nsicke}, {Kawaler}, {Clemens}, {Dunlap}, {Winget}, \& {Armstrong}}]{2015ApJ...810L...5H}
{Hermes}, J.~J., {Montgomery}, M.~H., {Bell}, K.~J., {et~al.} 2015, \apjl, 810, L5

\bibitem[{{Kawaler}(1988)}]{1988IAUS..123..329K}
{Kawaler}, S.~D. 1988, in IAU Symposium, Vol. 123, Advances in Helio- and Asteroseismology, ed. J.~{Christensen-Dalsgaard} \& S.~{Frandsen}, 329

\bibitem[{{Leggett} {et~al.}(2018){Leggett}, {Bergeron}, {Subasavage}, {Dahn}, {Harris}, {Munn}, {Ables}, {Canzian}, {Guetter}, {Henden}, {Levine}, {Luginbuhl}, {Monet}, {Monet}, {Pier}, {Stone}, {Vrba}, {Walker}, {Tilleman}, {Xu}, \& {Dufour}}]{2018ApJS..239...26L}
{Leggett}, S.~K., {Bergeron}, P., {Subasavage}, J.~P., {et~al.} 2018, \apjs, 239, 26

\bibitem[{{Liebert} {et~al.}(2005){Liebert}, {Bergeron}, \& {Holberg}}]{2005ApJS..156...47L}
{Liebert}, J., {Bergeron}, P., \& {Holberg}, J.~B. 2005, \apjs, 156, 47

\bibitem[{{Montgomery} {et~al.}(2003){Montgomery}, {Metcalfe}, \& {Winget}}]{2003MNRAS.344..657M}
{Montgomery}, M.~H., {Metcalfe}, T.~S., \& {Winget}, D.~E. 2003, \mnras, 344, 657

\bibitem[{{Munday} {et~al.}(2024){Munday}, {Pelisoli}, {Tremblay}, {Marsh}, {Nelemans}, {B{\'e}dard}, {Toonen}, {Breedt}, {Cunningham}, {O'Brien}, \& {Dawson}}]{Munday2024}
{Munday}, J., {Pelisoli}, I., {Tremblay}, P.~E., {et~al.} 2024, \mnras, 532, 2534

\bibitem[{{O'Brien} {et~al.}(2024){O'Brien}, {Tremblay}, {Klein}, {Koester}, {Melis}, {B{\'e}dard}, {Cukanovaite}, {Cunningham}, {Doyle}, {G{\"a}nsicke}, {Gentile Fusillo}, {Hollands}, {McCleery}, {Pelisoli}, {Toonen}, {Weinberger}, \& {Zuckerman}}]{2024MNRAS.527.8687O}
{O'Brien}, M.~W., {Tremblay}, P.~E., {Klein}, B.~L., {et~al.} 2024, \mnras, 527, 8687

\bibitem[{{O'Donoghue}(1994)}]{1994MNRAS.270..222O}
{O'Donoghue}, D. 1994, \mnras, 270, 222

\bibitem[{{Ricker} {et~al.}(2015){Ricker}, {Winn}, {Vanderspek}, {Latham}, {Bakos}, {Bean}, {Berta-Thompson}, {Brown}, {Buchhave}, {Butler}, {Butler}, {Chaplin}, {Charbonneau}, {Christensen-Dalsgaard}, {Clampin}, {Deming}, {Doty}, {De Lee}, {Dressing}, {Dunham}, {Endl}, {Fressin}, {Ge}, {Henning}, {Holman}, {Howard}, {Ida}, {Jenkins}, {Jernigan}, {Johnson}, {Kaltenegger}, {Kawai}, {Kjeldsen}, {Laughlin}, {Levine}, {Lin}, {Lissauer}, {MacQueen}, {Marcy}, {McCullough}, {Morton}, {Narita}, {Paegert}, {Palle}, {Pepe}, {Pepper}, {Quirrenbach}, {Rinehart}, {Sasselov}, {Sato}, {Seager}, {Sozzetti}, {Stassun}, {Sullivan}, {Szentgyorgyi}, {Torres}, {Udry}, \& {Villasenor}}]{2015JATIS...1a4003R}
{Ricker}, G.~R., {Winn}, J.~N., {Vanderspek}, R., {et~al.} 2015, Journal of Astronomical Telescopes, Instruments, and Systems, 1, 014003

\bibitem[{{Sahu} {et~al.}(2023){Sahu}, {G{\"a}nsicke}, {Tremblay}, {Koester}, {Hermes}, {Wilson}, {Toloza}, {Hoskin}, {Farihi}, {Manser}, \& {Redfield}}]{Sahu2023}
{Sahu}, S., {G{\"a}nsicke}, B.~T., {Tremblay}, P.-E., {et~al.} 2023, \mnras, 526, 5800

\bibitem[{{Saumon} {et~al.}(2022){Saumon}, {Blouin}, \& {Tremblay}}]{2022PhR...988....1S}
{Saumon}, D., {Blouin}, S., \& {Tremblay}, P.-E. 2022, \physrep, 988, 1

\bibitem[{{Tassoul} {et~al.}(1990){Tassoul}, {Fontaine}, \& {Winget}}]{1990ApJS...72..335T}
{Tassoul}, M., {Fontaine}, G., \& {Winget}, D.~E. 1990, \apjs, 72, 335

\bibitem[{{Tremblay} \& {Bergeron}(2009)}]{Tremblay2009}
{Tremblay}, P.~E. \& {Bergeron}, P. 2009, \apj, 696, 1755

\bibitem[{{Tremblay} {et~al.}(2015){Tremblay}, {Gianninas}, {Kilic}, {Ludwig}, {Steffen}, {Freytag}, \& {Hermes}}]{Tremblay2015}
{Tremblay}, P.~E., {Gianninas}, A., {Kilic}, M., {et~al.} 2015, \apj, 809, 148

\bibitem[{{Tremblay} {et~al.}(2013){Tremblay}, {Ludwig}, {Steffen}, \& {Freytag}}]{Tremblay2013}
{Tremblay}, P.~E., {Ludwig}, H.~G., {Steffen}, M., \& {Freytag}, B. 2013, \aap, 559, A104

\bibitem[{{Uzundag} {et~al.}(2023){Uzundag}, {De Ger{\'o}nimo}, {C{\'o}rsico}, {Silvotti}, {Bradley}, {Montgomery}, {Catelan}, {Toloza}, {Bell}, {Kepler}, {Althaus}, {Kleinman}, {Kilic}, {Mullally}, {G{\"a}nsicke}, {B{\k{a}}kowska}, {Barber}, \& {Nitta}}]{2023MNRAS.526.2846U}
{Uzundag}, M., {De Ger{\'o}nimo}, F.~C., {C{\'o}rsico}, A.~H., {et~al.} 2023, \mnras, 526, 2846

\bibitem[{{Uzundag} {et~al.}(2021){Uzundag}, {Vu{\v{c}}kovi{\'c}}, {N{\'e}meth}, {Miller Bertolami}, {Silvotti}, {Baran}, {Telting}, {Reed}, {Shoaf}, {{\O}stensen}, \& {Sahoo}}]{2021A&A...651A.121U}
{Uzundag}, M., {Vu{\v{c}}kovi{\'c}}, M., {N{\'e}meth}, P., {et~al.} 2021, \aap, 651, A121

\bibitem[{{Winget} \& {Kepler}(2008)}]{2008ARA&A..46..157W}
{Winget}, D.~E. \& {Kepler}, S.~O. 2008, \araa, 46, 157

\bibitem[{{Winget} {et~al.}(1991){Winget}, {Nather}, {Clemens}, {Provencal}, {Kleinman}, {Bradley}, {Wood}, {Claver}, {Frueh}, {Grauer}, {Hine}, {Hansen}, {Fontaine}, {Achilleos}, {Wickramasinghe}, {Marar}, {Seetha}, {Ashoka}, {O'Donoghue}, {Warner}, {Kurtz}, {Buckley}, {Brickhill}, {Vauclair}, {Dolez}, {Chevreton}, {Barstow}, {Solheim}, {Kanaan}, {Kepler}, {Henry}, \& {Kawaler}}]{1991ApJ...378..326W}
{Winget}, D.~E., {Nather}, R.~E., {Clemens}, J.~C., {et~al.} 1991, \apj, 378, 326

\bibitem[{{Winget} {et~al.}(1982){Winget}, {van Horn}, {Tassoul}, {Fontaine}, {Hansen}, \& {Carroll}}]{1982ApJ...252L..65W}
{Winget}, D.~E., {van Horn}, H.~M., {Tassoul}, M., {et~al.} 1982, \apjl, 252, L65

\bibitem[{{Zima}(2008)}]{2008CoAst.155...17Z}
{Zima}, W. 2008, Communications in Asteroseismology, 155, 17

\end{thebibliography}

\begin{appendix}

\begin{table*}
\section{Frequency, period, and amplitude values determined from the TESS data sets.}
\label{app:A}
\centering
\caption{Frequency, period, and amplitude values determined from the TESS data sets.}
\label{tabl:freqs}
\begin{tabular}{lrrrrr}
\hline
\hline
\multicolumn{1}{c}{} & \multicolumn{1}{c}{$f\,[\mu$Hz]} & \multicolumn{1}{c}{P [s]} & \multicolumn{1}{c}{A [mma]} & \multicolumn{1}{c}{Comments} & \multicolumn{1}{c}{Lin. comb.}\\
\hline
$f_{01}$ & 634.0 & 1577.3 & 6.6 & \citet{2018MNRAS.478.2676B} & \\
$f_{02}$ & 668.0 & 1497.0 & 1.08 & only in s22 & \\
$f_{03}$ & 717.4 & 1394.0 & 1.07 & s22 & \\
$f_{04}$ & 927.3 & 1078.4* & 0.98 & s4849 & \\
$f_{05}$ & 963.0 & 1038.4* & 7.0 & \citet{2018MNRAS.478.2676B} & \\
$f_{06}$ & 997.2 & 1002.8 & 1.70 & s22 & \\
$f_{07}$ & 1003.0 & 997.1* & 2.41 & s22 & \\
$f_{08}$ & 1012.6 & 987.5 & 1.12 & s22 & \\
$f_{09}$ & 1017.8 & 982.5 & 1.44 & s22 & \\
$f_{10}$ & 1032.7 & 968.4 & 2.98 &  & \\
$f_{11}$ & 1038.3 & 963.1 & 2.66 & s22 & \\
$f_{12}$ & 1041.5 & 960.1 & 1.06 & s4849 & \\
$f_{13}$ & 1044.8 & 957.2 & 1.41 &  & \\
$f_{14}$ & 1048.6 & 953.6* & 1.95 &  & \\
$f_{15}$ & 1076.9 & 928.6 & 3.69 &  & \\
$f_{16}$ & 1082.6 & 923.7 & 1.48 & s7576 & \\
$f_{17}$ & 1087.8 & 919.3* & 2.08 & s7576 & \\
$f_{18}$ & 1095.9 & 912.5 & 0.85 & s4849 & \\
$f_{19}$ & 1091.6 & 916.1 & 1.14 & s7576 & \\
$f_{20}$ & 1126.9 & 887.4 & 2.37 & s7576 & \\
$f_{21}$ & 1144.5 & 873.8* & 1.12 & s7576 & \\
$f_{22}$ & 1149.5 & 869.9 & 0.91 & s7576 & \\
$f_{23}$ & 1183.9 & 844.7 & 1.31 & s7576 & \\
$f_{24}$ & 1193.9 & 837.6* & 10.45 & s7576 & \\
$f_{25}$ & 1250.3 & 799.8 & 2.67 & s7576 & \\
$f_{26}$ & 1260.3 & 793.4* & 3.94 & s7576 & \\
$f_{27}$ & 1389.1 & 719.9 & 0.86 & s1516 & \\
$f_{28}$ & 1435.0 & 696.9 & 0.97 & s1516 & \\
$f_{29}$ & 1438.9 & 695.0 & 4.74 &  & \\
$f_{30}$ & 1456.0 & 686.8 & 1.76 & s1516 & \\
$f_{31}$ & 1452.6 & 688.4 & 1.16 & s4849 & \\
$f_{32}$ & 1478.9 & 676.2* & 1.08 &  & \\
$f_{33}$ & 1601.3 & 624.5 & 0.82 & s4849 & \\
$f_{34}$ & 1727.9 & 578.7 & 1.05 & s22 & \\
$f_{35}$ & 1751.9 & 570.8 & 1.87 &  & \\
$f_{36}$ & 1844.2 & 542.2 & 0.78 & s4849 & \\
$f_{37}$ & 1969.4 & 507.8* & 2.94 &  & \\
$f_{38}$ & 1996.0 & 501.0 & 1.21 &  & \\
$f_{39}$ & 2558.3 & 390.9* & -- & \citet{2018MNRAS.473.3693G} & \\
$f_{40}$ & 2739.0 & 365.1 & 2.39 &  & \\
$f_{41}$ & 2759.0 & 362.4 & 1.40 &  & \\
\hline
 & 362.2 & 2761.2 & 1.01 & s1516 & yes (s1516; $f_{29} - f_{15}$) \\
 & 404.8 & 2470.2 & 0.98 &  & yes (s1516, s22; $f_{29} - f_{10}$) \\
 & 775.2 & 1290.0 & 1.38 & s7576 & yes (s7576; $f_{37} - f_{24}$)\\
 & 967.5 & 1033.6 & 1.07 & s22 & yes (s22; $f_{37} - f_{07}$)\\ 
 & 1544.9 & 647.3 & 1.04 & s7576 & yes (s7576; $f_{40} - f_{24}$)\\
  & 1720.8 & 581.1 & 1.89 &  & yes (s22; $f_{40} - f_{09}$)\\
 & 2071.0 & 482.9 & 1.32 &  & yes (s22; $f_{10} + f_{11}$)\\
 & 2183.9 & 457.9 & 1.29 &  & yes (s1516; half-int. $1.5*f_{30}$)\\
 & 2387.7 & 418.8 & 0.85 & s7576 & yes (s7576; harmonic $2*f_{24}$)\\  
\hline
\end{tabular}
\tablefoot{Only three periods are from the literature, which are indicated in the fourth column. Where similar frequencies were found in multiple TESS data sets, their amplitude-averaged period is listed in the table. Periods that form a regular pattern of $\ell= 1$ modes with constant period spacing are marked with an asterisk (see Sect. \ref{ps_tests}). In cases where a significant frequency was found in only one data set, the sector number is indicated in the fourth column. The fifth column notes if the given frequency was identified as a linear combination in any data set.}
\end{table*}

\FloatBarrier
\twocolumn

\begin{table}[h!]
\section{Frequencies, periods and amplitudes of the detected signals in the data sets of the different sectors.}
\label{app:B}
\caption {Results of the analysis of the sector s15s16 data set.}
\label{table:s15s16} 
\centering
\begin{tabular}{lrrr}
\hline
\hline
\multicolumn{1}{c}{} & \multicolumn{1}{c}{$f\,[\mu$Hz]} & \multicolumn{1}{c}{P [s]} & \multicolumn{1}{c}{A [mma]} \\
\hline
F19 & 362.2 & 2761.2 & 1.0 \\
F32 & 403.2 & 2479.9$^+$ & 0.8 \\
F31 & 404.1 & 2474.6$^+$ & 0.9 \\
F30 & 405.5 & 2466.0 & 0.8 \\
F17 & 1032.6 & 968.5$^+$ & 1.4 \\
F13 & 1033.2 & 967.9 & 1.3 \\
F18 & 1034.3 & 966.8$^+$ & 1.1 \\
F03 & 1076.2 & 929.2$^+$ & 2.3 \\
F10 & 1076.5 & 929.0$^+$ & 1.8 \\
F02 & 1076.7 & 928.7 & 3.2 \\
F09 & 1077.1 & 928.5$^+$ & 2.1 \\
F15 & 1078.4 & 927.3$^+$ & 1.5 \\
F20 & 1079.3 & 926.5$^+$ & 1.1 \\
F25 & 1389.1 & 719.9 & 0.9 \\
F21 & 1435.0 & 696.9 & 1.0 \\
F26 & 1438.1 & 695.4$^+$ & 0.9 \\
F01 & 1438.9 & 695.0 & 5.8 \\
F23 & 1439.2 & 694.8$^+$ & 1.0 \\
F29 & 1454.3 & 687.6$^+$ & 1.1 \\
F08 & 1455.1 & 687.2$^+$ & 2.2 \\
F33 & 1455.7 & 686.9$^+$ & 1.0 \\
F07 & 1456.0 & 686.8 & 1.8 \\
F27 & 1456.5 & 686.6$^+$ & 1.0 \\
F28 & 1457.7 & 686.0$^+$ & 0.9 \\
F24 & 1478.3 & 676.4$^+$ & 1.0 \\
F14 & 1479.2 & 676.0 & 1.0 \\
F12 & 1721.3 & 580.9 & 1.4 \\
F05 & 1751.9 & 570.8 & 2.6 \\
F04 & 1969.7 & 507.7 & 2.7 \\
F16 & 1995.9 & 501.0 & 1.2 \\
F22 & 2184.0 & 457.9 & 1.0 \\
F06 & 2739.2 & 365.1 & 2.5 \\
F11 & 2759.1 & 362.4 & 1.5 \\
\hline
\end{tabular}
\tablefoot{We mark with a plus sign the frequencies close to a larger amplitude peak. In these cases the amplitude determinations are not completely reliable. We did not consider these frequencies as unique ones in the next steps of our frequency analysis.}
\end{table}

\begin{table}[h!]
\caption {Results of the analysis of the sector s22 data set.}
\label{table:s22} 
\centering
\begin{tabular}{lrrr}
\hline
\hline
\multicolumn{1}{c}{} & \multicolumn{1}{c}{$f\,[\mu$Hz]} & \multicolumn{1}{c}{P [s]} & \multicolumn{1}{c}{A [mma]} \\
\hline
F31 & 406.3 & 2461.5 & 1.1 \\
F36 & 668.0 & 1497.0 & 1.1 \\
F38 & 717.4 & 1394.0 & 1.1 \\
F37 & 967.5 & 1033.6 & 1.1 \\
F33 & 996.1 & 1003.9$^+$ & 1.2 \\
F13 & 997.2 & 1002.8 & 1.7 \\
F18 & 1002.0 & 998.0$^+$ & 1.6 \\
F07 & 1003.0 & 997.1 & 2.4 \\
F09 & 1003.8 & 996.2$^+$ & 2.1 \\
F34 & 1012.6 & 987.5 & 1.1 \\
F21 & 1017.8 & 982.5 & 1.4 \\
F23 & 1030.6 & 970.3$^+$ & 1.3 \\
F02 & 1032.6 & 968.4 & 5.2 \\
F22 & 1033.7 & 967.4$^+$ & 1.3 \\
F12 & 1037.4 & 964.0 $^+$& 1.7 \\
F04 & 1038.3 & 963.1 & 2.7 \\
F11 & 1039.3 & 962.2$^+$ & 1.9 \\
F32 & 1040.3 & 961.2$^+$ & 1.2 \\
F26 & 1043.2 & 958.6$^+$ & 1.3 \\
F35 & 1044.0 & 957.9$^+$ & 1.2 \\
F16 & 1045.1 & 956.8 & 1.7 \\
F17 & 1046.2 & 955.9$^+$ & 1.5 \\
F27 & 1047.1 & 955.0$^+$ & 1.3 \\
F08 & 1048.1 & 954.1 & 2.3 \\
F14 & 1049.2 & 953.1$^+$ & 1.7 \\
F15 & 1050.2 & 952.2$^+$ & 1.6 \\
F25 & 1051.2 & 951.3$^+$ & 1.4 \\
F19 & 1075.8 & 929.6$^+$ & 1.5 \\
F03 & 1076.8 & 928.7 & 3.6 \\
F20 & 1077.8 & 927.8$^+$ & 1.5 \\
F01 & 1438.9 & 695.0 & 7.2 \\
F29 & 1721.1 & 581.0 & 1.2 \\
F39 & 1727.9 & 578.7 & 1.1 \\
F10 & 1751.7 & 570.9 & 2.0 \\
F05 & 1969.6 & 507.7 & 2.5 \\
F30 & 1995.6 & 501.1 & 1.1 \\
F24 & 2071.0 & 482.9 & 1.3 \\
F40 & 2072.0 & 482.6$^+$ & 1.0 \\
F28 & 2183.8 & 457.9 & 1.2 \\
F06 & 2739.0 & 365.1 & 2.4 \\
\hline
\end{tabular}
\end{table}

\begin{table}[h!]
\caption {Results of the analysis of the sector s48s49 data set.}
\label{table:s48s49} 
\centering
\begin{tabular}{lrrr}
\hline
\hline
\multicolumn{1}{c}{} & \multicolumn{1}{c}{$f\,[\mu$Hz]} & \multicolumn{1}{c}{P [s]} & \multicolumn{1}{c}{A [mma]} \\
\hline
F34 & 402.7 & 2483.2 & 1.0 \\
F35 & 927.3 & 1078.4 & 1.0 \\
F52 & 1031.4 & 969.6$^+$ & 0.8 \\
F08 & 1032.6 & 968.5 & 2.4 \\
F14 & 1032.7 & 968.3$^+$ & 1.6 \\
F24 & 1033.7 & 967.4$^+$ & 1.2 \\
F33 & 1034.2 & 966.9$^+$ & 1.1 \\
F49 & 1040.0 & 961.5$^+$ & 0.9 \\
F50 & 1041.2 & 960.4$^+$ & 1.0 \\
F48 & 1041.5 & 960.1 & 1.1 \\
F36 & 1044.2 & 957.6 & 1.2 \\
F51 & 1044.5 & 957.4$^+$ & 0.9 \\
F18 & 1046.4 & 955.6$^+$ & 1.7 \\
F19 & 1047.7 & 954.5$^+$ & 1.6 \\
F46 & 1048.1 & 954.1$^+$ & 1.0 \\
F39 & 1049.2 & 953.1$^+$ & 1.2 \\
F15 & 1049.3 & 953.0 & 1.6 \\
F41 & 1049.7 & 952.6$^+$ & 1.2 \\
F26 & 1050.3 & 952.1$^+$ & 1.2 \\
F17 & 1050.8 & 951.6$^+$ & 1.2 \\
F23 & 1052.0 & 950.6$^+$ & 1.6 \\
F38 & 1052.4 & 950.2$^+$ & 1.2 \\
F30 & 1053.8 & 948.9$^+$ & 1.2 \\
F27 & 1055.1 & 947.8$^+$ & 1.2 \\
F16 & 1074.8 & 930.4$^+$ & 1.7 \\
F10 & 1075.7 & 929.6$^+$ & 3.1 \\
F02 & 1076.0 & 929.3$^+$ & 3.8 \\
F03 & 1076.6 & 928.8$^+$ & 4.2 \\
F01 & 1077.2 & 928.3 & 4.2 \\
F07 & 1077.4 & 928.2$^+$ & 3.1 \\
F43 & 1078.3 & 927.4$^+$ & 1.0 \\
F11 & 1078.5 & 927.2$^+$ & 2.1 \\
F42 & 1095.9 & 912.5 & 0.9 \\
F44 & 1127.7 & 886.8$^+$ & 0.9 \\
F21 & 1128.7 & 886.0$^+$ & not valid \\
F12 & 1128.7 & 885.9$^+$ & not valid \\
F06 & 1129.2 & 885.6 & 2.7 \\
F32 & 1129.4 & 885.4$^+$ & 1.5 \\
F47 & 1438.3 & 695.3$^+$ & 0.8 \\
F25 & 1438.8 & 695.0 & 1.2 \\
F37 & 1452.6 & 688.4 & 1.2 \\
F54 & 1453.3 & 688.1$^+$ & 0.9 \\
F55 & 1457.8 & 686.0$^+$ & 0.8 \\
F31 & 1478.1 & 676.5 & 1.1 \\
F28 & 1478.6 & 676.3 & 1.1 \\
F45 & 1601.3 & 624.5 & 0.8 \\
F04 & 1720.8 & 581.1 & 3.0 \\
F40 & 1751.7 & 570.9$^+$ & 1.0 \\
F13 & 1751.9 & 570.8 & 1.7 \\
F53 & 1844.2 & 542.2 & 0.8 \\
F05 & 1969.3 & 507.8 & 2.6 \\
F29 & 1995.9 & 501.0 & 1.1 \\
F22 & 2183.9 & 457.9 & 1.3 \\
F09 & 2738.9 & 365.1 & 2.3 \\
F20 & 2758.9 & 362.5 & 1.4 \\
\hline
\end{tabular}
\tablefoot{Not valid amplitudes mean that we detect two frequencies so close to each other that we cannot resolve them correctly.}
\end{table}

\begin{table}[h!]
\caption {Results of the analysis of the sector s75s76 data set.}
\label{table:s75s76} 
\centering
\begin{tabular}{lrrr}
\hline
\hline
\multicolumn{1}{c}{} & \multicolumn{1}{c}{$f\,[\mu$Hz]} & \multicolumn{1}{c}{P [s]} & \multicolumn{1}{c}{A [mma]} \\
\hline
F34 & 1082.2 & 924.1$^+$ & 1.2 \\
F15 & 1082.6 & 923.7 & 1.5 \\
F39 & 1085.2 & 921.5$^+$ & 1.1 \\
F19 & 1087.1 & 919.9$^+$ & 1.7 \\
F18 & 1087.8 & 919.3 & 2.1 \\
F42 & 1091.1 & 916.5$^+$ & 1.0 \\
F28 & 1091.6 & 916.1 & 1.1 \\
F06 & 1126.8 & 887.4 & not valid \\
F09 & 1126.8 & 887.4$^+$ & not valid \\
F24 & 1127.2 & 887.2$^+$ & 1.6 \\
F48 & 1129.8 & 885.1$^+$ & 1.0 \\
F36 & 1131.0 & 884.2 & 1.1 \\
F35 & 1133.6 & 882.2$^+$ & 1.2 \\
F32 & 1133.9 & 881.9 & 1.3 \\
F43 & 1144.5 & 873.8 & 1.1 \\
F46 & 1145.0 & 873.4$^+$ & 1.1 \\
F49 & 1149.5 & 869.9 & 0.9 \\
F22 & 1183.9 & 844.7 & 1.3 \\
F31 & 1184.4 & 844.3$^+$ & 1.6 \\
F38 & 1185.0 & 843.9$^+$ & 1.4 \\
F37 & 1186.0 & 843.2$^+$ & 1.3 \\
F45 & 1188.1 & 841.7$^+$ & 1.1 \\
F20 & 1190.9 & 839.7$^+$ & 1.5 \\
F51 & 1192.9 & 838.3$^+$ & 1.2 \\
F03 & 1193.7 & 837.7$^+$ & 8.6 \\
F01 & 1193.9 & 837.6 & 10.4 \\
F02 & 1194.2 & 837.4$^+$ & 7.9 \\
F10 & 1195.4 & 836.6$^+$ & 2.4 \\
F44 & 1195.5 & 836.4$^+$ & 1.3 \\
F07 & 1196.6 & 835.7 & 2.2 \\
F16 & 1197.1 & 835.3$^+$ & not valid \\
F13 & 1197.1 & 835.3$^+$ & not valid \\
F41 & 1245.2 & 803.1$^+$ & 1.0 \\
F12 & 1250.0 & 800.0$^+$ & 2.1 \\
F11 & 1250.3 & 799.8 & 2.7 \\
F27 & 1251.7 & 798.9$^+$ & 1.4 \\
F21 & 1260.3 & 793.4$^+$ & not valid \\
F04 & 1260.3 & 793.4 & not valid \\
F50 & 1261.7 & 792.6$^+$ & 1.0 \\
F29 & 1263.0 & 791.8$^+$ & 1.4 \\
F23 & 1263.6 & 791.4$^+$ & 1.8 \\
F47 & 1264.1 & 791.1$^+$ & 1.0 \\
F40 & 1544.9 & 647.3 & 1.0 \\
F14 & 1720.3 & 581.3 & 1.9 \\
F33 & 1752.5 & 570.6 & 1.2 \\
F05 & 1969.2 & 507.8 & 3.9 \\
F25 & 1996.4 & 500.9 & 1.4 \\
F17 & 2184.0 & 457.9 & 1.7 \\
F52 & 2387.7 & 418.8 & 0.9 \\
F08 & 2739.0 & 365.1 & 2.3 \\
F30 & 2759.0 & 362.5 & 1.3 \\
F26 & 775.2 & 1290.0 & 1.4 \\
\hline
\end{tabular}
\end{table}

\end{appendix}

\end{document}